\documentclass{elsarticle}

\usepackage{lineno,hyperref}
\modulolinenumbers[5]

\journal{Journal of Computational Physics}

\usepackage{amssymb}

\graphicspath{{./img/}}

\usepackage{amssymb}
\usepackage{amsthm}
\usepackage{lineno}

\usepackage{float}
\usepackage{multirow}
\usepackage{booktabs}
\usepackage[dvipsnames]{xcolor} 
\usepackage{array}
\usepackage{xspace}   	   
\usepackage{mathrsfs}
\usepackage{mdwlist}         
\usepackage{placeins}
\usepackage[english]{babel}

\usepackage[linesnumbered,lined,boxed,commentsnumbered]{algorithm2e}

\usepackage{bigints}
\usepackage{tensor}
\usepackage{multicol}
\usepackage{subfig}
\usepackage{tikz}
\usetikzlibrary{calc}
\usetikzlibrary{patterns}
\usetikzlibrary{arrows}
\usetikzlibrary{arrows.meta}
\usetikzlibrary{shadings}
\usetikzlibrary{positioning}
\usepackage{pgfplots}
\usepackage{listings}

\usepackage[normalem]{ulem}

\usepackage{dsfont}
\newcommand{\reals}{\ensuremath{\mathds{R}}}

\newcommand{\inputtikzfile}[1]{%
\IfFileExists{./tikz/#1.pdf}{\includegraphics[scale=1]{./tikz/#1.pdf}}{\input{tikz/#1.tikz}}%
}

\newcommand{\x}{\ensuremath{\mathbf{x}}}

\makeatletter
\DeclareFontFamily{U}{tipa}{}
\DeclareFontShape{U}{tipa}{m}{n}{<->tipa10}{}
\newcommand{\arc@char}{{\usefont{U}{tipa}{m}{n}\symbol{62}}}%

\newcommand{\arc}[1]{\mathpalette\arc@arc{#1}}

\newcommand{\arc@arc}[2]{%
  \sbox0{$\m@th#1#2$}%
  \vbox{
    \hbox{\resizebox{\wd0}{\height}{\arc@char}}
    \nointerlineskip
    \box0
  }%
}
\makeatother

\definecolor{liquid}{rgb}{0.28,0.46,1}

\makeatletter
\newcommand\resetsubfigs{\setcounter{sub\@captype}{0}}
\makeatother

\begin{document}

\begin{frontmatter}

  \title{Automatic surface mesh generation for discrete models ~--~ A complete and automatic pipeline based on reparametrization}

\author[a,b]{Pierre-Alexandre Beaufort\corref{mycorrespondingauthor}}
\author[b]{Christophe Geuzaine}
\author[a]{Jean-Fran\c cois Remacle}

\address[a]{Universit\'e catholique de Louvain, iMMC, Avenue Georges Lemaitre 4, 1348 Louvain-la-Neuve, Belgium}
\address[b]{Universit\'e de Li\`ege, Montefiore Institute, All\'ee de la D\'ecouverte 10, B-4000 Li\`ege, Belgium}

\cortext[mycorrespondingauthor]{Corresponding author}

\begin{abstract}
  Triangulations are an ubiquitous input for the finite element community.
  However, most raw triangulations obtained by imaging techniques are unsuitable
  as-is for finite element analysis.  In this paper, we give a robust pipeline
  for handling those triangulations, based on the computation of a one-to-one
  parametrization for automatically selected patches of input triangles, which
  makes each patch amenable to remeshing by standard finite element meshing
  algorithms.  Using only geometrical arguments, we prove that a discrete
  parametrization of a patch is one-to-one if (and only if) its image in the
  parameter space is such that all parametric triangles have a positive area. We
  then derive a non-standard linear discretization scheme based on mean value
  coordinates to compute such one-to-one parametrizations, and show that the
  scheme does \emph{not} discretize a Laplacian on a structured mesh. The
  proposed pipeline is implemented in the open source mesh generator Gmsh, where
  the creation of suitable patches is based on triangulation topology and
  parametrization quality, combined with feature edge detection. Several
  examples illustrate the robustness of the resulting implementation.
\end{abstract}


\end{frontmatter}

\section{Introduction}

Engineering designs are often encapsulated in Computer Aided Design (CAD)
systems. This is usually the case in automotive, shipbuilding or aerospace
industries.  The finite element method is the proeminent technique for
performing analysis of these designs and this method requires a finite element
mesh, i.e. a subdivision of CAD geometrical entities into a (large) collection
of simple geometrical shapes such as triangles, quadrangles, tetrahedra and
hexahedra, arranged in such a way that if two of them intersect, they do so
along a face, an edge or a node, and never otherwise.

In CAD systems, the geometry of surfaces is described through a parametrization
i.e. a mapping
\begin{equation}
\label{eqn:param}
\x~:~A \mapsto \reals^3,~~~(u;v) \mapsto \x (u;v)
\end{equation}
where $A\subset \reals^2$ is usually a rectangular region
$[u_0, u_1] \times [v_0, v_1]$. When finite element mesh generation procedures
have access to such parametrizations $\x (u;v)$ of surfaces, it is in general a
good idea to generate a planar mesh in the parametric domain $A$ and map it in
3D. This way of generating surface meshes is called \emph{indirect} (\cite{borouchaki2000parametric}),
and is the predominant method for generating high-quality finite element surface
meshes in a robust manner. This approach is in particular followed by the open
source mesh generator Gmsh (\cite{geuzaine2009gmsh}), which directly interacts
with CAD systems to get parametrizations $\x (u;v)$ as well as their derivatives
(normals, curvatures...). The nature of the mappings $\x (u;v)$ that are
provided by CAD systems is such that anisotropic planar meshing capabilities are
required in order to be able to generate quality meshes in 3D. Those mappings
may be very irregular and even singular, for example at the two poles of a
sphere. Gmsh's surface planar mesh generators have been designed in such a way
that they can handle very distorted metrics (\cite{remacle2019gmsh}) while still
providing high quality outputs.

In domains like geophysics or in bio-sciences, however, the geometry of the
models is rarely described through CAD models. Most often, those geometries are
produced through imaging (segmentation, voxelization, ...) whose geometrical
output can be reduced to a triangulation. Several authors have proposed
\emph{direct} approaches (such as \cite{frey2001yams,bechet2002generation}) that modify
this raw ``geometrical'' mesh to produce a ``computational'' mesh with elements
of controlled shapes and sizes. The aim of this paper is to show that the
\emph{indirect} approach is also possible in this case, in which a (global)
parametrization $\x (u;v)$ is not readily available. Starting from a
triangulation, our aim is to build \emph{a set of parametrizations} that form an
atlas of the model, and which can be used as-is by existing finite element mesh
generators.

This paper describes the complete pipeline that allows to build the atlas of the
model together with the parametrizations of all its maps. It aims at being
self-consistent, which makes it quite exhaustive. In \S\ref{sec:ma}, some
theoretical background on mappings is presented. Then, \S\ref{sec:discreteparam}
develops the concept of discrete parametrizations. A complete set of proofs
based on purely geometrical arguments is given that assert the injectivity of
the discrete maps that are used.  The way Gmsh handles the input in order to
ease the parametrization and meshing process is described within \S
\ref{sec:pipeline}.  We point out the drawback of a general processing of coarse
discrete surfaces in \S\ref{sec:coarse}, and discuss two ways to handle such
coarse discretizations. Several examples are presented in 
\S\ref{sec:examples}, and conclusions are drawn in 
\S\ref{sec:conclusion}.

\FloatBarrier
\section{Mappings}
\label{sec:ma}

A parametrization $\x(u;v) $ as defined in Equation \eqref{eqn:param}
is regular if $\partial_u  \x$ and $\partial_v  \x$
exist and are linearly independent:
$$\partial_u  \x  \times \partial_v  \x\neq {\mathbf 0}$$
for any $(u;v) \in A$.  In other words, $\x(u;v) $ is regular if
and only if the Jacobian matrix
\begin{equation}
\label{eqn:jac}
J =\dfrac{\partial \x}{\partial (u;v)} \in \reals^{3 \times 2}
\end{equation}
 associated to $\x(u;v)$ has rank 2 $\forall (u;v) \in A.$
The nature of the mapping $\x(u;v)$ is fully characterized by the
\emph{singular value decomposition} (SVD) of its Jacobian
\eqref{eqn:jac}. Its singular values $\sigma_1 \geq \sigma_2 > 0$
allow to characterize $\x$:
\begin{itemize}
\item $\x$ is \emph{isometric} if and only if $\sigma_1 = \sigma_2 = 1$,
\item $\x$ is \emph{conformal} if and only if $\dfrac{\sigma_2}{\sigma_1}=1$,
\item $\x$ is \emph{equiareal} if and only if $\sigma_1 \sigma_2 = 1$.
\end{itemize}

Isometric parametrizations preserve essentially everything (lengths, areas and
angles). With such nice properties, generating well shaped triangles in the
planar domain $A$ will lead to a well shaped mesh in 3D. Disappointingly, such
length preserving mappings do not exist for surfaces that are not developable
\citep[Chapter 2, §4]{struik1961lectures} i.e. that have non zero Gaussian
curvatures.

Conformal mappings conserve angles. If $\x(u;v)$ is conformal, isotropy is
preserved and standard isotropic mesh generators can do the surface meshing
job. Again, the odds are against us: although it is possible to build conformal
mappings for most surfaces, it is very difficult to ensure global injectivity of
such mappings, even though conformal mappings are always locally
injective. Thus, ensuring the global one-to-oneness of conformal mappings is
still an open question (see \cite{levy2002least}).

\begin{figure}[ht!]
\begin{center}
\subfloat[Mesh in the parameter plane $A$.]{\includegraphics[width=.475\linewidth]{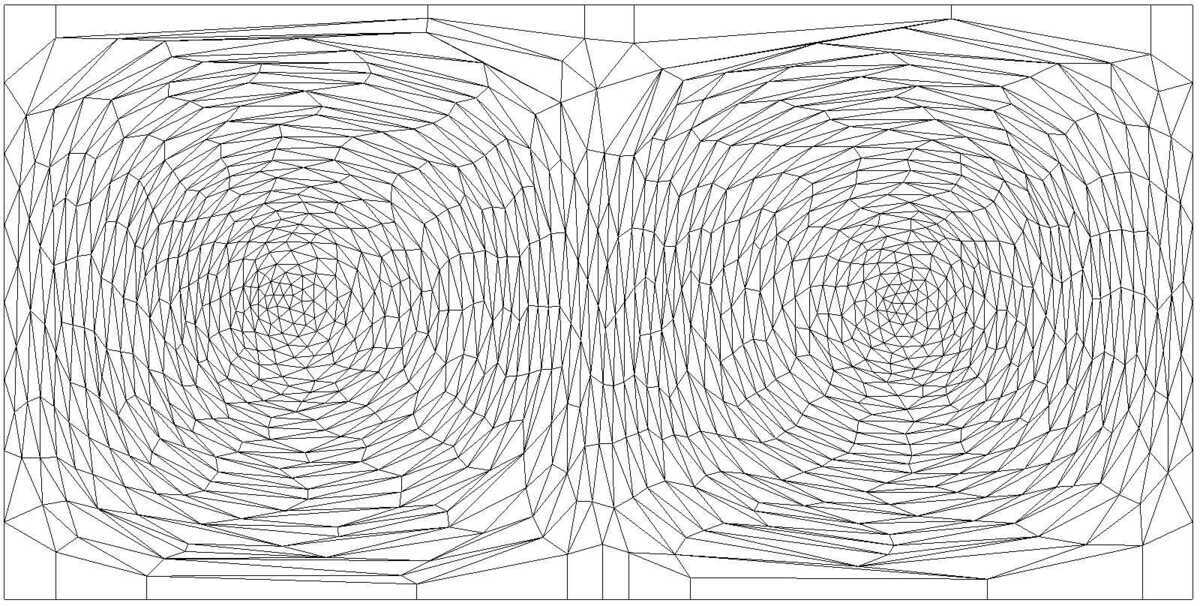}\label{sub:ellipsoidParam}}\hspace*{.25cm}
\subfloat[Mesh in $\reals^3$.]{\includegraphics[width=.475\linewidth]{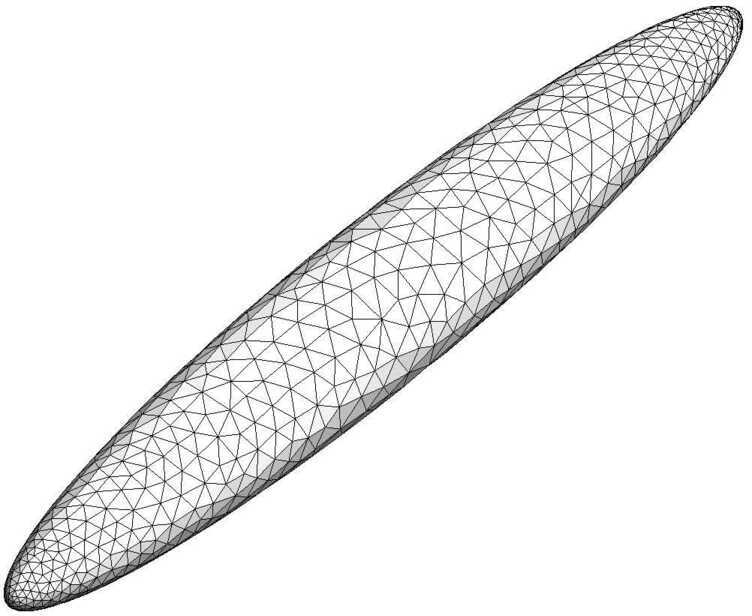}}\\
\subfloat[Largest singular value $\sigma_1$ in $A$.]{\includegraphics[width=.475\linewidth]{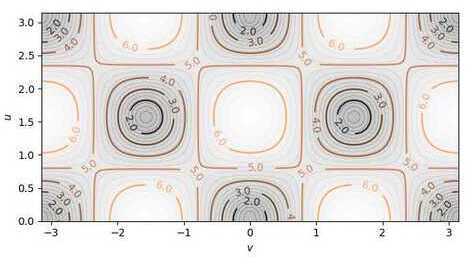}\label{sub:ellipsoidSigma1}}\hspace*{.25cm}\subfloat[Smallest singular value $\sigma_2$ in $A$.]{\includegraphics[width=.475\linewidth]{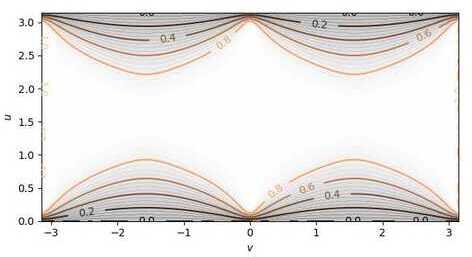}\label{sub:ellipsoidSigma2}}\\
\subfloat[Conformity $\frac{\sigma_2}{\sigma_1}$ in $A$.]{\includegraphics[scale=.5]{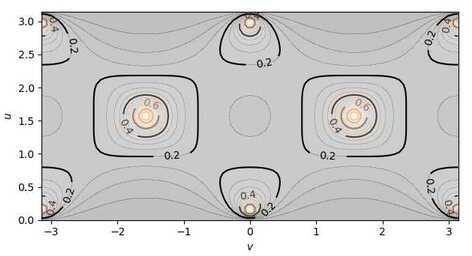}\label{sub:ellipsoidConformity}}
\end{center}
\caption{The case of an ellipsoid. Fig. a shows the mesh of the
  ellipsoid in the parameter space while Fig. b shows the same mesh
  in the 3D space. Fig. c and d show the largest and smallest
  singular values $\sigma_1$ and $\sigma_2$ of the jacobian
  $J$. Fig. e shows the non conformity parameter $\sigma_2/\sigma_1$.  }
\label{fig:ellipsoid}
\end{figure}

Equiareal mappings have no interest in mesh generation. Thus, in general, mesh
generators are faced with general parametrizations that do not preserve
anything. This means that anisotropic planar mesh generators are required to
generate well shaped meshes in 3D. Anisotropic mesh generators usually take as
input a Riemannian metric field defined in each $(u;v)$ of $A$. If the aim is to
produce an isotropic 3D mesh with a mesh size defined by an isotropic mesh size
field $h(\x(u;v))$, the metric tensor that is used by the mesh generator will be
$$
M(u;v) = {J^T J \over h^2}.
$$
Let us assume for example that the surface to be meshed is an
ellipsoid. Fig.~\ref{fig:ellipsoid} shows a 3D surface mesh that is adapted to
the maximal curvature of the surface as well as its counterpart in the parameter
plane of the ellipsoid. The particular ellipsoid of Fig.~\ref{fig:ellipsoid} is
$e = 7$ times wider in the $x$ direction than in the two other directions $y$
and $z$. Its parametrization (which is standard to most CAD systems) is
$$
{\displaystyle {\begin{aligned}x(u,v)&= e \sin u \,\sin v \\y(u,v)&=\sin u
      \,\cos v \\z(u,v)&=\cos u \end{aligned}}}
$$
where $u \in [0,\pi]$ is the inclination and $v \in [-\pi,\pi[$ is the azimuth.
The metric tensor associated to that mapping is
\begin{equation}
{\small
M = {1 \over h^2}
\begin{pmatrix}
\cos^2v (e^2 \sin^2v + \cos^2v) + \sin^2u &
\sin u \sin v \cos u \cos v (e^2-1) \\
\sin u \sin v \cos u \cos v (e^2-1) &
\sin^2u (e^2 \cos^2v + \sin^2v)
\end{pmatrix}
\label{eqn:m}
}
\end{equation}
The mapping is obviously not regular when $u=0$ and when $u=\pi$.  This is
surprisingly not so much of a problem for mesh generators. \cite{remacle2019gmsh}
propose a way to slightly modify meshing procedures in order to deal
with singular mappings such as the one of the ellipsoid. The metric field
\eqref{eqn:m} is anisotropic (see Fig.~\ref{sub:ellipsoidConformity}) and
non-uniform.  Yet it is smooth and smoothness of mappings is the most important
feature of $\x(u;v)$ in order to allow mesh generators to do a good job.  When
generating a mesh in an indirect fashion, a planar mesh, possibly anisotropic,
is generated in the parameter plane $A$. Then, one may think that this planar
mesh is mapped in 3D through $\x(u;v)$, which is not true: only corners of the
triangles are mapped in 3D and those corners are connected together with 3D
straight lines that are not the actual mapping of 2D straight lines. In the best
case scenario, any 2D straight line connecting points $(u_a;v_a)$ and
$(u_b;v_b)$ corresponds the geodesic between those two points. When the metric
$M$ is locally constant, geodesics are straight lines and the indirect meshing
approach gives good results. When the metric varies rapidly along one given
edge, then indirect meshing becomes difficult. In CAD systems, parametrizations
are always smooth and indirect mesh generation is always possible.

\FloatBarrier
\section{Discrete Parametrizations}
\label{sec:discreteparam}

In Section \S\ref{sec:ma}, we have shown that having a smooth
parametrization was the condition to allow indirect surface meshing.
CAD systems provide smooth parametrizations but CAD models are
not the only geometrical representations that are available in
engineering analysis. In many domains of engineering interest,
geometries of models are described by triangulations. We call such models
\emph{discrete models}.

Assume a triangulation $T$ with $\#p$ nodes (vertices), $\#e$ edges and $\#t$
triangles which are correctly oriented to each others. Finding a parametrization of $T$ consists in assigning to
every vertex $p_i$ of the triangulation a pair of coordinates
$(u_i;v_i)$. If every triangle $(p_i,p_j,p_k)$, with $p_\bullet\in \reals^3$\footnote{In what follows, a triangle is denoted by the indices of its nodes, i.e. $(i,j,k)$ instead of $(p_i,p_j,p_k)$.} of the triangulation
has a positive area in the $(u;v)$ plane, then the parametrization is
injective.

A parametrization of $T$ onto a subset of $A \subset \reals^2$
exists if the triangulation corresponds to the one
of a planar mesh. Assume that triangulation $T$ is simply connected
with $\#b$ boundaries, $\#h$ vertices on those boundaries and whose the genus is $g$.
Then the surface is parameterizable if and only if
$$\#t = 2 (\#p -1) + 2(\#b -1) - \#h + 4g$$

In what follows, we present some existing material that is detailed in numerous
publications such as \cite{floater2005surface,tutte1963draw,remacle2010high}. The main interest of this section is that we take
here the point of view of the numerical geometer. The main result about the
one-to-oneness of mappings is proven without using one single theorem of
analysis such as maximum principles of Radó-Kneser-Choquet
theorem (see \citep{Dirac1953888}). 

Consider an internal vertex $i$ of $T$ and $J(i)$ the set of indices whose the corresponding nodes are connected to the node $i$ (in other words, edge $(i,j)$ exists $\forall j \in J(i)$).
The value of the parametric coordinates $(u_i,v_i)$ at vertex $i$ will be computed as a weighted average of the coordinates $(u_{j},v_{j})$ of its neighboring vertices:
\begin{equation}
\sum_{j \in J(i)}  \lambda_{ij} (u_i -u_{j}) = 0~~,~~
\sum_{j \in J(i)}  \lambda_{ij} (v_i -v_{j}) = 0
\label{eqn:exp}
\end{equation}
where $\lambda_{ij}$ are coefficients. This scheme is a called a
\emph{difference scheme}
that involves only differences $(u_i-u_{j})$, with $j \in J(i)$.
If every $\lambda_{ij}$ is positive, values of $u_i$ and $v_i$ are convex
combinations of their surrounding values. In a geometrical point of
view, it actually means that point $(u_i,v_i)$ lies in the convex hull
${\mathcal H}_i$ of its
neighboring vertices.

\begin{figure}[!ht]
\begin{center}
\subfloat[Stencil around vertex $\mathbf{i}$.]{\includegraphics[width=.45\linewidth]{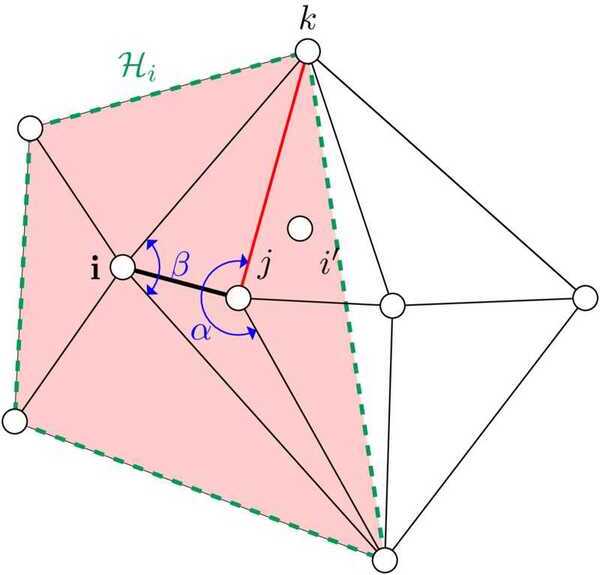}\label{sub:inside}}\hspace*{.25cm}
\subfloat[Stencil around vertex $\mathbf{j}$ with $i'$.]{\includegraphics[width=.45\linewidth]{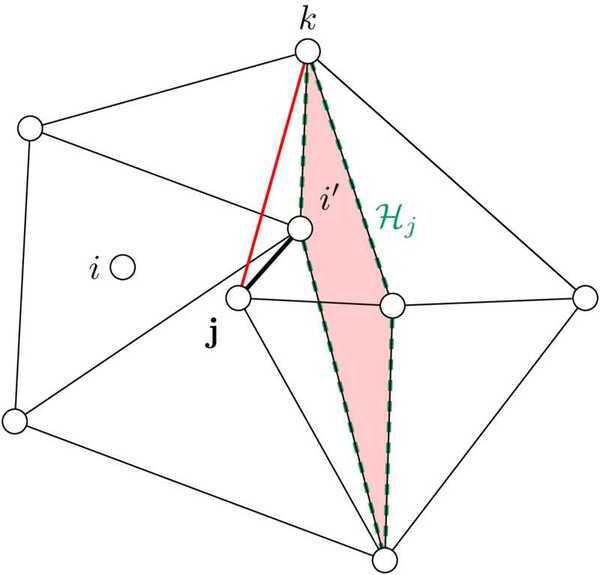}\label{sub:outside}}
\end{center}
\caption{Sketch of proof for monotonicity.}
\label{fig:hull}
\end{figure}

With that assumption, it is easy to prove that the mapping
provided by any positive scheme of the type \eqref{eqn:exp} is
one-to-one. Let us consider a triangle $(i,j,k)$ in the parameter
plane $(u;v)$, Fig.\ref{fig:hull}.
If edge $(j,k)$ belongs to ${\mathcal H}_i$, that triangle $(i,j,k)$ is
obviously positive.

On the other hand, if $(j,k)$ is inside ${\mathcal H}_i$,
as it is the case in Fig. \ref{sub:inside}, then $(j,k)$ does
not belong to  ${\mathcal H}_i$ and moving $i$ to $i'$
creates an inverted triangle $(i',j,k)$ while keeping
${\mathcal H}_i = {\mathcal H}_{i'}$. In this case, $i'$
is inside  ${\mathcal H}_{i}$ while triangle $(i',j,k)$
is inverted. It is easy to see that moving $i$ to $i'$ implies
that $j$ would be outside ${\mathcal H}_{j}$ which is in
contradiction with the hypothesis that each vertex is inside its
convex hull, Fig.\ref{sub:outside}. Vertex $j$ being inside  ${\mathcal H}_{i}$ implies
that $\alpha > \pi$. The sum of the four angles of a quadrangle
is $2 \pi$. This implies that $\beta < \pi$ which implies
that edge $(i,k)$ belongs to  ${\mathcal H}_{j}$. So, moving $i$ to
$i'$ puts $j$ outside  ${\mathcal H}_{j}$.

Now see what happens on the outer boundary $\partial A$
of the $(u;v)$ domain $A$. There, points have no neighboring hull.
Yet, assuming that $\partial A$ is convex, then all vertices of
$\partial A$ that are connected to internal vertices belong to the
convex hull of those latter internal vertices.
Besides, no internal vertex cannot be situated outside $A$.
It means that all triangles having (at least) one edge belonging to $\partial A$ are positive.
This last part of the proof has some similarities with the one
of \cite{floater2003mean}.

This means that a positive scheme applied to a convex domain implies
that the discrete parametrization is one-to-one. In our case, we will
always choose $\partial A$ as the unit circle.

Now, the right choice of the $\lambda_{ij}$ is of outmost importance  for
ensuring a good parametrization. Our use of parametrization is
meshing. The first and non negotiable property of the discrete
parametrization is one-to-oneness. We thus choose a positive scheme
and a $(u;v)$ domain that is a unit circle.
The second priority is smoothness, we will develop that aspect below.
The icing on the cake would be conformity (i.e. angle preservation) but, as noted in
\S\ref{sec:ma}, Gmsh's mesh generators
are comfortable with anisotropic mappings and we will not put any
effort on that aspect of the game (our aim is not texture mapping like
in computer graphics, so we are OK to map squares on circles).

\FloatBarrier
\subsection{Parametrization smoothness}
We look here for a smooth function $\x(u,v)$ i.e. a
continuous function whose derivatives are smooth as well
because we want $\sigma_1$ and $\sigma_2$ to be smooth and
$\sigma_1$ and $\sigma_2$  are by-products of the metric i.e.
a tensor computed using the first derivatives of $\x(u,v)$.
Tutte’s  barycentric  mapping (see \cite{tutte1963draw}) consists in choosing
$\lambda_{ij}=1$. This choice leads to very irregular mappings
that are useless for mesh generation purposes. The idea that has been
advocated by many authors (e.g. \cite{levy2002least,marchandise2011high}) is to
solve a partial differential equation whose solutions are inherently
smooth. For example, the solution of Laplace equations on domains with
smooth boundaries and with smooth boundary conditions are
$C^{\infty}$ and it is indeed a good idea to choose the $\lambda_{ij}$
in such a way the difference operator \eqref{eqn:exp} is a discrete
version of the Laplace operator.

\FloatBarrier
\subsection{Laplace smoothing using $\mathcal{P}^1$  finite elements}

\begin{figure}[!ht]
\begin{center}
\includegraphics[width=.3\linewidth]{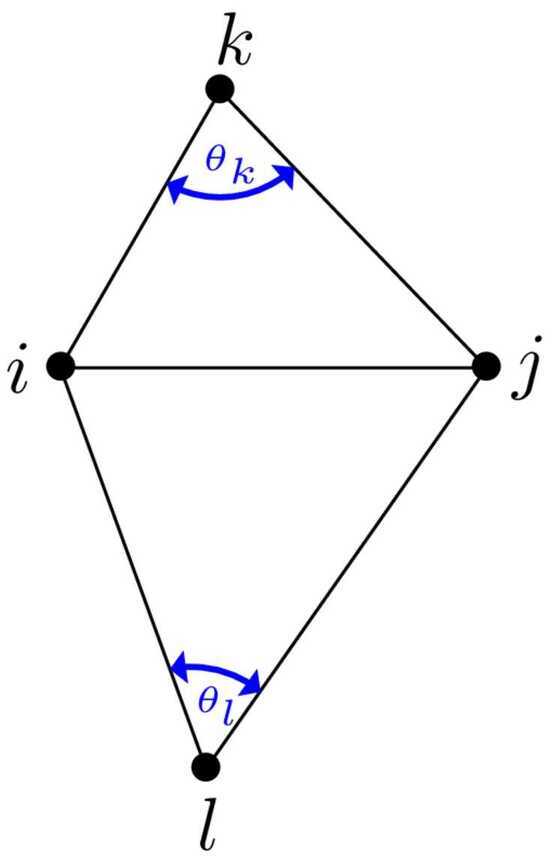}
\caption{Definitions of $\theta_k$ and $\theta_l$ for the difference scheme corresponding to the linear Galerkin approach.}
\label{fig:thetafem}
\end{center}
\end{figure}

The standard $P^1$ finite element formulation of the Laplace problem
is well known for more than a half of a century. In the early days,
some authors (\cite{duffin1959distributed}) have written coefficients
$\lambda_{ij}^{\text{FEM}}$ in a quite geometrical fashion (see Fig. \ref{fig:thetafem}):
\begin{equation}\label{eqn:fem}
\lambda_{ij}^{\text{FEM}} := \dfrac{1}{2} \left( \dfrac{\cos(\theta_k)}{\sin(\theta_k)} + \dfrac{\cos(\theta_l)}{\sin(\theta_l)} \right).
\end{equation}
For sake of completeness, the so called ``cotangent formula'' \eqref{eqn:fem} is
fully derived in~\ref{appendixFEM}.  Coefficients $\lambda_{ij}^{\text{FEM}}$ of
\eqref{eqn:fem} may be negative for
$\theta_{\bullet} \in \left(\frac{\pi}{2};\pi\right)$, which could lead to
scheme that is not provably injective; \cite[\S 5]{floater1998parametric} gives a simple example where the Laplacian smoothing fails to provide an injective mapping.
This is the very old result that states
that the maximum principle satisfied by solutions of Laplace equations is only
guaranteed \emph{a priori} by finite elements computed on acute triangulations,
i.e. triangulations without obtuse angles.
Acute triangulations are a sufficient condition for injectivity. Yet it is not
necessary and it is indeed complicated to find examples where finite elements
fail to provide one-to-one parametrizations. Disappointingly, in the world of
mesh generation, limit cases that happen once in a thousand have to be avoided.
So, we will not use the finite element version of Laplacian smoothing for parametrizing our surfaces.

\FloatBarrier
\subsection{Mean value coordinates}
\label{sec:mvc}

\begin{figure}[!ht]
\begin{center}
\subfloat[Contribution of a triangle.]{\includegraphics[height=.4\linewidth]{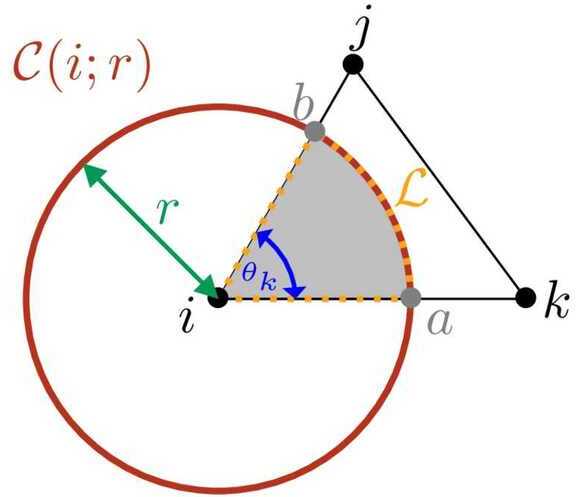}\label{sub:sketchmvc}}\hspace*{.25cm}
\subfloat[Definitions of $\theta_k$ and $\theta_l$.]{\includegraphics[height=.4\linewidth]{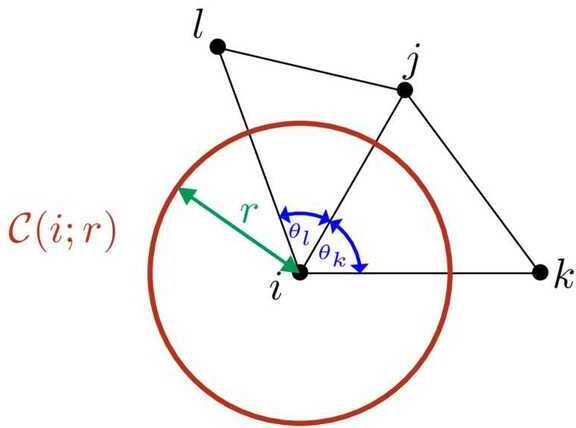}\label{sub:thetamvc}}
\caption{Derivation of the difference scheme corresponding to the mean value coordinates.}
\label{fig:mvcderivation}
\end{center}
\end{figure}

A continuous function $f$ is a solution of Laplace equation $\nabla^2 f = 0$ on
an open set $A \subset \reals^2$ if and only if, for every $\x \in A$, $f(\x)$
is equal to the average value of $f$ over every circle of radius $r$
$\mathcal{C}(\mathbf{x};r)$ that fully belongs to $A$:
\begin{equation}\label{eqn:maxmin}
f(\mathbf{x}) = \dfrac{1}{2 \pi r}
\int_{\mathcal{C}(\mathbf{x};r)} f(\x')~d\x'.
\end{equation}
This principle states that the extrema of the mapping are located on the boundary of the domain, and that there is not local extremum inside the domain.

\citep{floater2003mean} proposes a way to compute $\lambda_{ij}$
that actually mimics property \eqref{eqn:maxmin}: this scheme is called mean
value coordinates. In this paper, we re-derive Floater's $\lambda_{ij}$
corresponding to mean value coordinates using a finite element point of view.
According to \eqref{eqn:maxmin}, the value $f_i$ is the average of values
$f(\mathbf{x})$ along a circle $\mathcal{C}(i;r)$ of radius $r$ centered on $i$
(see Fig. \ref{fig:mvcderivation}).  A linear interpolation
$f(x;y)=\sum_j f_j \phi_j(x;y)$ is assumed over each triangle
$\mathcal{T}_{ijk}$.  We are going to compute the contribution of triangle
$\mathcal{T}_{ijk}$ for \eqref{eqn:maxmin}
$$
\theta_k r ~ f_i = \int_{\arc{ab}} f_i \phi_i + f_j \phi_j  + f_k \phi_k ~ ds
$$
where $\theta_k$ is the angle between edges $[ij]$ and $[ik]$, and $\arc{ab}$ is the circle arc of $\partial\mathcal{C}(i,r)$ contained in $\mathcal{T}_{ijk}$, Fig. \ref{sub:sketchmvc}.
Since $\phi_i + \phi_j + \phi_k = 1$,
$$
\underbrace{\left(\theta_k r - \int_{\arc{ab}} \phi_i ~ ds\right)}_{\int_{\arc{ab}} \phi_j + \phi_k ~ ds} f_i - \int_{\arc{ab}} \phi_j ~ ds f_j - \int_{\arc{ab}}  \phi_k ~  ds f_k = 0
$$
which gives
$$
\underbrace{\int_{\arc{ab}} \phi_j ~ ds}_{\lambda_{ij}} (f_i-f_j) +  \underbrace{\int_{\arc{ab}} \phi_k ~ ds}_{\lambda_{ik}} (f_i-f_k) = 0
$$
over $\mathcal{T}_{ijk}$.

Linear shape function $\phi_j$ associated to node $j$ in $\mathcal{T}_{ijk}$ corresponds to
$$
\phi_j(x;y) =  \dfrac{y}{y_j}
$$
where $y$ is the vertical coordinate relative to edge $[ik]$ and $y_j$ is the $y$-coordinate of node $j$.
We compute the integral of $y$ over $\arc{ab}$ from the contour $\mathcal{C}$ composed of $\arc{ab}$, $[bi]$ and $[ia]$
$$
\int_{\mathcal{C}} y ~ ds
$$

From normal vector of $\mathcal{C}(i;r)$ $\hat{n} = \frac{1}{r}(x;y)$, we get
$$
\int_{\mathcal{L}} \dfrac{y}{r} ~ ds = \int_{\mathcal{L}} \hat{n} \cdot \mathbf{e_y} ~ ds
$$
with $\mathbf{e_y}=(0;1)$.
The divergence of $\mathbf{e_y}$ is obviously zero, and owing to the divergence theorem
$$
\int_{\mathcal{L}} \dfrac{y}{r} ~ ds = \int_{\mathcal{R}(\mathcal{L})} \nabla \cdot \mathbf{e_y} ~dx~dy = 0
$$
where $\mathcal{R}(\mathcal{L})$ is the region surrounded by $\mathcal{L}$ (gray area, Fig. \ref{sub:sketchmvc}).
The integral along the circle arc $\arc{ab}$ is then equal to the opposite of integrals along edges $[bi]$ and $[ia]$ of triangles $\mathcal{T}_{ijk}$
$$
\begin{array}{rcl}
\displaystyle{\int_{\arc{ab}} \underbrace{\hat{n} \cdot \mathbf{e_y}}_{\dfrac{y}{r}} ~ ds} &=& \displaystyle{- \left( \int_{[ia]} \underbrace{\hat{n} \cdot \mathbf{e_y}}_{-1} ~ds + \int_{[bi]} \underbrace{\hat{n} \cdot \mathbf{e_y}}_{\cos(\theta_k)} ~ds \right)} \\
&=& - (-r + r\cos(\theta_k))\\
&=& r (\cos(\theta_k)-1)
\end{array}
$$
Since $y_j = l_{ij} \sin(\theta_k)$, with $l_{ik}$ the length of edge $[ij]$
$$
\int_{\arc{ab}} \phi_k ~ ds = r^2 \dfrac{\tan\left(\frac{\theta_k}{2}\right)}{l_{ij}}
$$

Choosing a radius $r$ small enough (i.e. smaller than the smallest edge within
the triangulation) allows to simplify the finite scheme (\ref{eqn:fem}) by
$r^2$, which means that the scheme does not depend on the circle of
integration. The coefficient $\lambda_{ij}$ is then given by
\begin{equation}\label{eqn:mvc}
\lambda_{ij} = \dfrac{\tan\left(\frac{\theta_k}{2}\right)+\tan\left(\frac{\theta_l}{2}\right)}{l_{ij}}
\end{equation}

We notice that $\lambda_{ij}>0,~\forall \theta_{\bullet} \in (0;\pi)$.  The
difference scheme (\ref{eqn:exp}) with (\ref{eqn:mvc}) builds linear injective
mappings.  This monotone scheme is not symmetric, except on equilateral
triangulations.

\begin{figure}[ht!]
\begin{center}
\subfloat[Structured.]{\includegraphics[width = .25\linewidth]{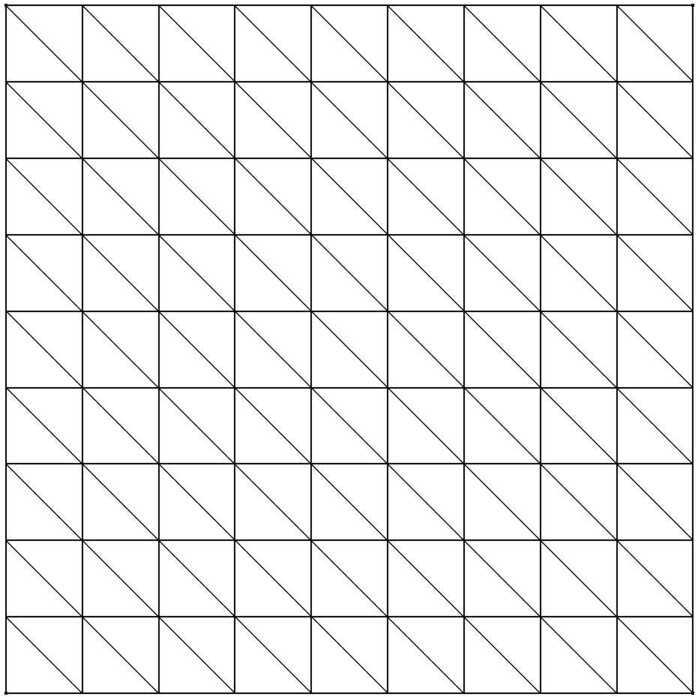}\label{sub:structured}}\hspace*{.25cm}
\subfloat[Delaunay.]{\includegraphics[width = .25\linewidth]{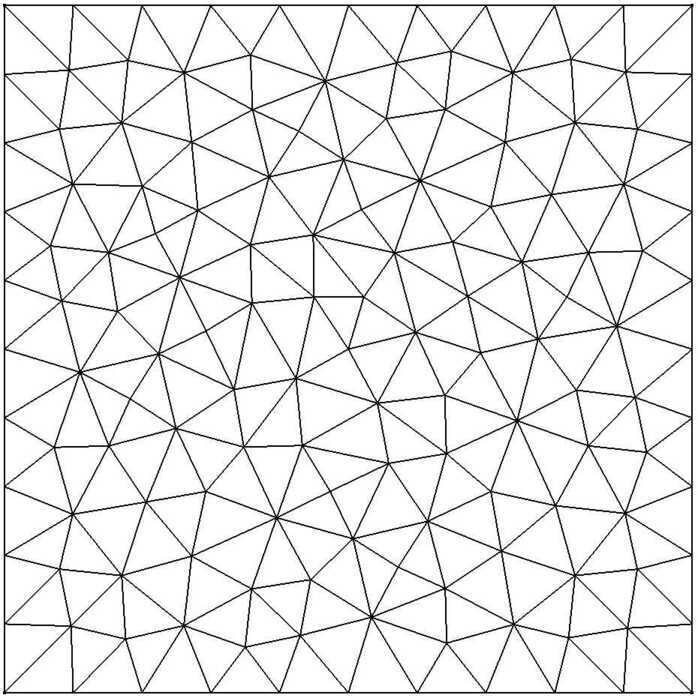}\label{sub:delaunay}}\hspace*{.25cm}
\subfloat[Frontal.]{\includegraphics[width = .25\linewidth]{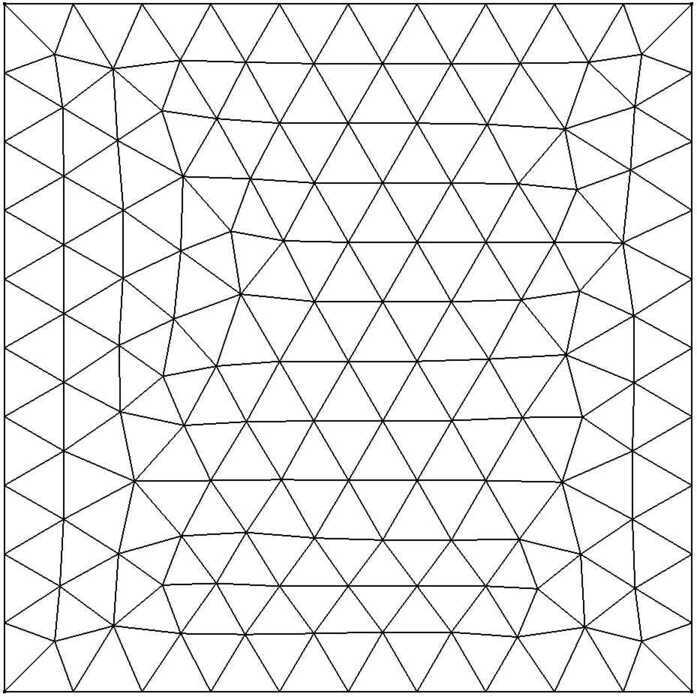}\label{sub:frontal}}
\end{center}
\caption{Types of meshes on a square.}
\label{fig:meshes}
\end{figure}

\begin{figure}[ht!]
\begin{center}
\subfloat[$L^2$ norm.]{\includegraphics[width = .475\linewidth]{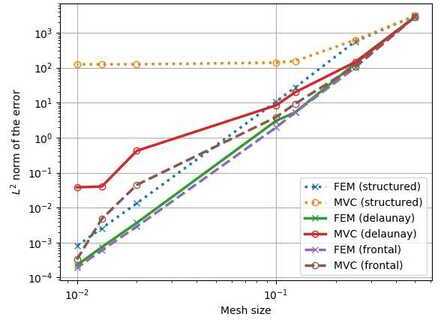}}\hspace*{.25cm}
\subfloat[$H^1$ seminorm.]{\includegraphics[width = .475\linewidth]{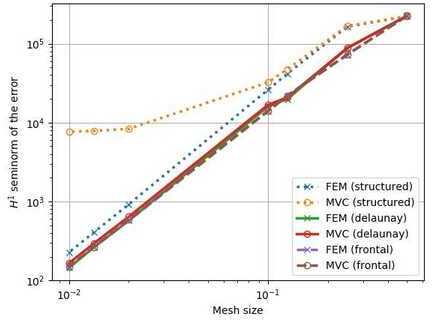}}
\end{center}
\caption{$h$-convergence of discrete schemes (\ref{eqn:exp}) with
  (\ref{eqn:fem}) VS (\ref{eqn:mvc}) on mesh types of Fig. \ref{fig:meshes}.}
\label{fig:convergence}
\end{figure}

At that point, one can raise the question of the actual accuracy of the MVC
scheme for discretizing Laplace equation, which is our guarantee of smoothness.
A convergence experiment\footnote{The experiment has been performed with the
  Gmsh API, given in supplementary material.} has been performed on a square
$[0;1]\times[0;1]$ on various meshes (Fig. \ref{fig:meshes}) using the
standard technique of manufactured solutions. We choose
$f(x;y) = \sin(2\pi x)\cosh(2\pi y)$ whose laplacian $\nabla^2 f$ is zero.

Fig. \ref{fig:convergence} shows that MVC scheme does not exhibit the usual
FEM convergence. The absence of symmetry of the MVC scheme implies that only
$\mathcal{O}(h)$ convergence is observed for the $L^2$ norm. Yet, the MVC scheme
seems to converge on all meshes except the structured one.
This behavior is due to the fact that the MVC scheme does not correspond to a Laplacian over a structured triangulation, \ref{appendixMVCvsLaplace}.

\FloatBarrier
\subsection{Boundary conditions}\label{sec:bcs}

\begin{figure}[ht!]
\begin{center}
\includegraphics[width=.65\linewidth]{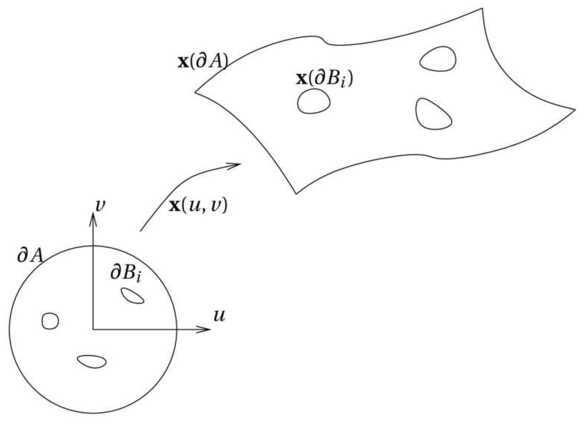}
\end{center}
\caption{ A 3D domain that is topologically equivalent to a disk with
$3$ internal boundaries and its parametric domain $A$.
\label{fig:BC}}
\end{figure}

We consider 3D surfaces that are topologically equivalent to a disk with
$\#b -1$ internal boundaries. The parametric domain that is considered is always
a unit disk
$$
A = \left\{(u;v)\in \Re^2: u^2+v^2 < 1\right\}.
$$
The setup is described in Fig. \ref{fig:BC}.
Dirichlet boundary conditions are applied on $\x(\partial A)$ that actually
ensure that the $u,v$ coordinates on $\x(\partial A)$ correspond to the unit
circle
$$
\delta A = \left\{(u;v)\in \Re^2: u^2+v^2 = 1\right\}.
$$
We should now decide on what boundary conditions to apply on the other
boundaries $\delta B_i$. The issue here is that we do not know \emph{a priori}
their position in the parameter plane. We could decide their position and insert
$\#b-1$ small circles inside $A$. Yet, this would lead to a parametrization that
is quite distorted. Another option is to apply the smoother as is to every
internal points, including the ones on the internal boundaries. This indeed
corresponds to imposing homogeneous Neumann boundary conditions on every
internal boundary.  It is indeed easy to prove that this choice still leads to a
one-to-one parametrization.  One first thing to note is that if every
$\partial B_i$ is convex and if we use a convex combination map like
\eqref{eqn:mvc}, then the mapping is one-to-one.

\begin{figure}[ht!]
\begin{center}
\includegraphics[width=.35\linewidth]{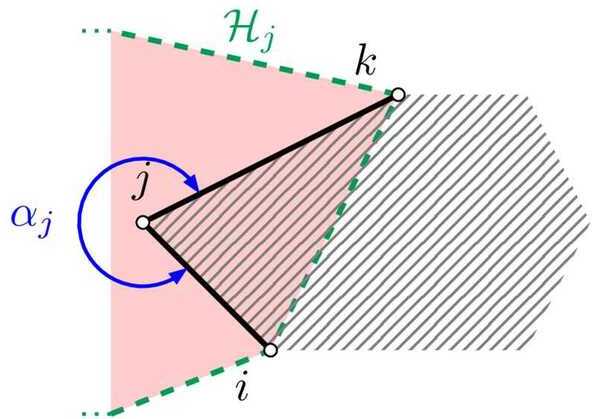}
\end{center}
\caption{Three consecutive points belonging to a loop describing an hole (hatched area) in $A$.}
\label{fig:convexhole}
\end{figure}

Assume that points $p_1,p_2,\dots,p_k$ form a closed loop in the parameter plane
and that every point lies in the convex hull of its neighbors, such as
Fig. \ref{fig:convexhole}. Then, polygon $(p_1,p_2,\dots,p_k)$ is convex.
Indeed, if every three consecutive points $i,j,k$ of such a loop form an angle
$\alpha_j$ that is greater or equal to $\pi$, then the edges $(i;j)$ and $(j;k)$
lay in the convex hull $\mathcal{H}_j$.  If it is true for every point of the
loop corresponding to the hole, then its loop in the parameter plane is convex.
From §\ref{sec:discreteparam}, we know that a positive scheme produce a
one-to-one parametrization.  Hence, if no condition are imposed on the holes -
which corresponds to Neumann condition within FEM formulation - the parametric
representation of those holes correspond to convex loop, whatever the initial
shape of holes (i.e. even if they were concave).

\begin{figure}[ht!]
\begin{center}
  \subfloat[Homogeneous Neumann.]{
    \begin{minipage}[12cm]{.4\linewidth}
      $\begin{array}{c}
\includegraphics[width = \linewidth]{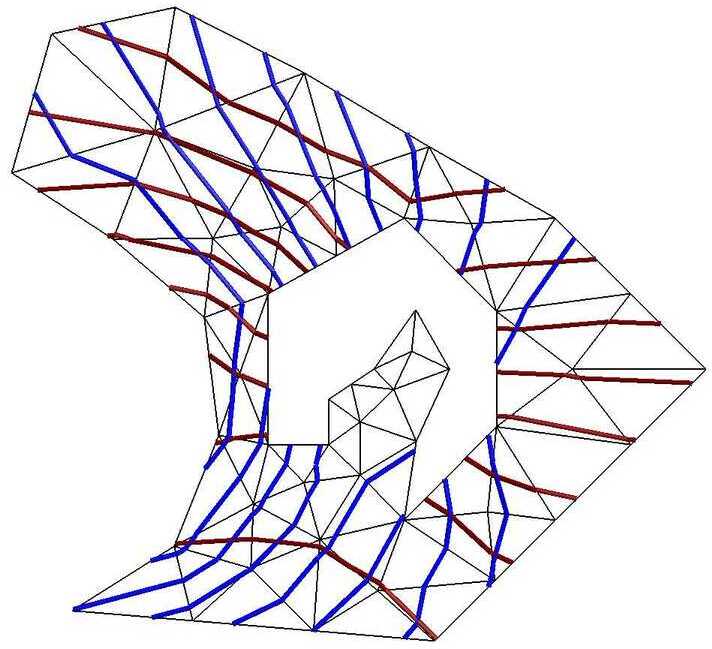}\\
         \begin{tikzpicture}
           \node at (0,0) {\includegraphics[width = \linewidth]{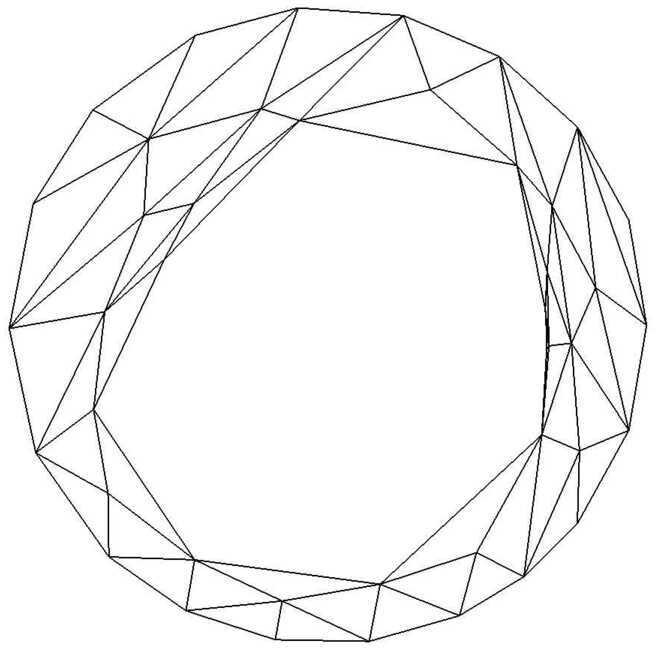}};
           \draw[ForestGreen,line width=.5mm] (1.5,-1) rectangle (1.8,.4);
         \end{tikzpicture}\\
         \begin{tikzpicture}[every node/.style={inner sep=0,outer sep=0}]
           \node[draw,line width=1mm,ForestGreen] at (0,0) {\includegraphics[scale = .15]{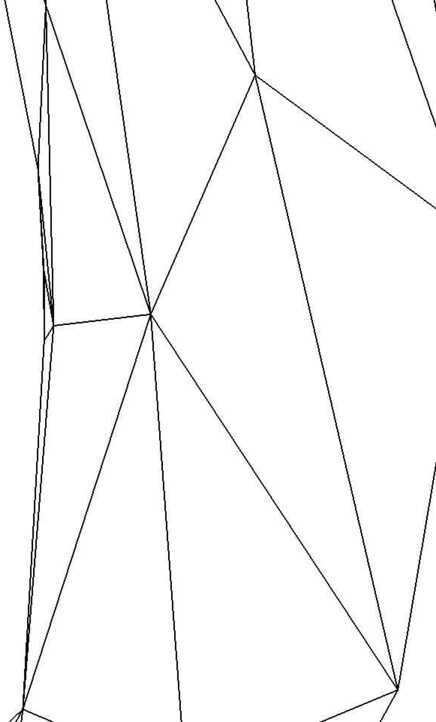}};
         \end{tikzpicture}
       \end{array}$
     \end{minipage}\label{sub:concaveNeumann}}\hspace*{.25cm}
   \subfloat[Filling hole.]{
     \begin{minipage}[12cm]{.4\linewidth}
       $\begin{array}{c}
          \includegraphics[width = \linewidth]{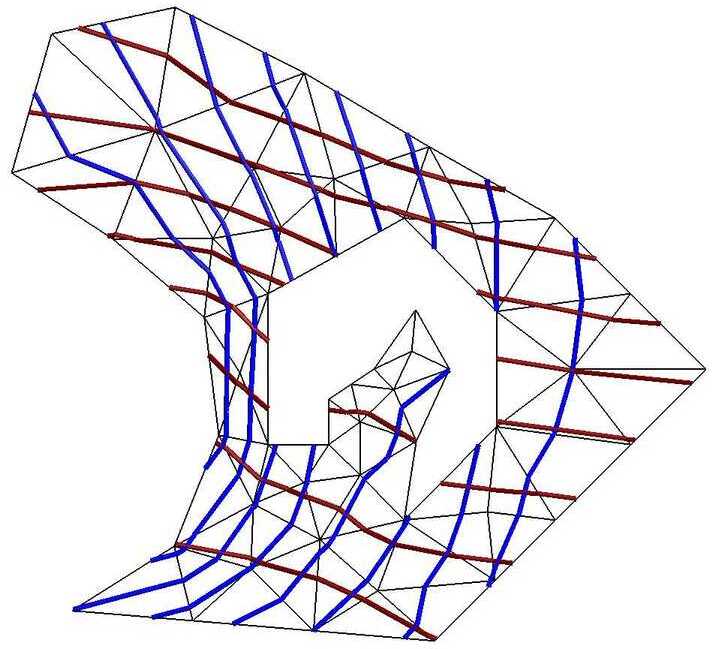}\\
          \begin{tikzpicture}
            \node at (0,0) {\includegraphics[width = \linewidth]{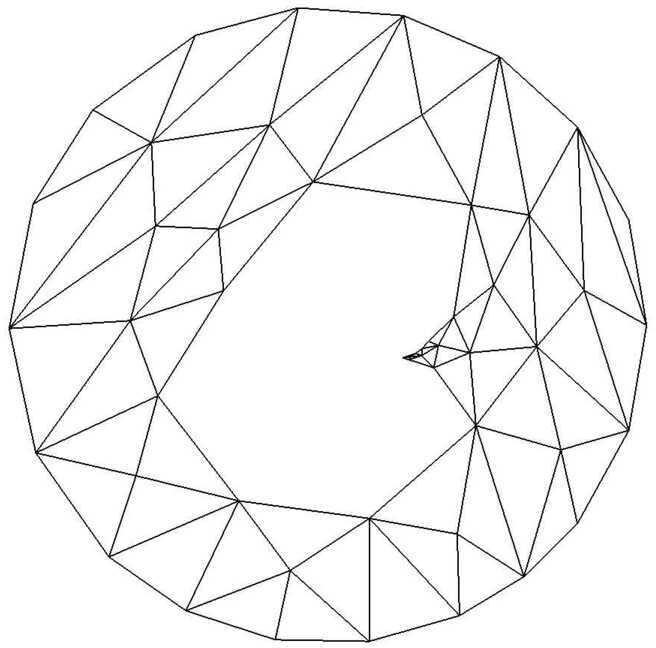}};
            \draw[ForestGreen,line width=.5mm] (.525,-.5) rectangle (1.3,.25);
          \end{tikzpicture}\\
          \begin{tikzpicture}[every node/.style={inner sep=0,outer sep=0}]
            \node[draw,line width=1mm,ForestGreen] at (0,0) {\includegraphics[width = \linewidth]{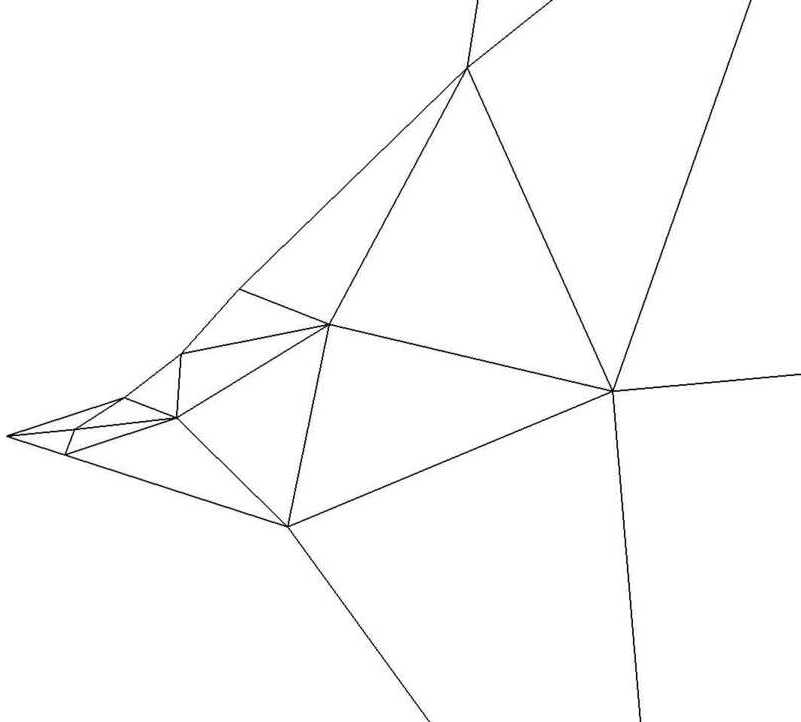}};
          \end{tikzpicture}
        \end{array}$
      \end{minipage}\label{sub:concaveFilling}
}\\
\end{center}
\caption{Demonstration of filling a concave hole with the circle
  assumption. Top: parametrization over the (discrete) geometry ($u$: red
  isolines, $v$: blue isolines). Bottom: triangles within the computed
  parametric space.}
\label{fig:concave}
\end{figure}

Fig. \ref{sub:concaveNeumann} shows a concave domain with a concave hole that
is mapped using \eqref{eqn:mvc} and where homogeneous Neuman boundary conditions
were applied to the internal boundary.  In this case,
$\partial_n u = \partial_n v = 0$ on the internal boundary and the
parametrization is close to be singular because the two tangent vectors are
nearly parallel: both of them are \emph{weakly} orthogonal to the boundary (see
Fig. \ref{sub:concaveNeumann})!

Another option consists of \emph{filling the holes}, which leads to better
results in practice (see Fig. \ref{sub:concaveFilling}).

\begin{figure}[!ht]
\begin{center}
\includegraphics[width=.65\linewidth]{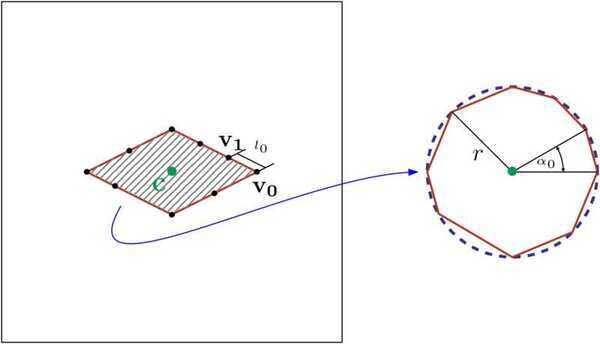}
\caption{Exampled filling hole (hatched area).}
\label{fig:filling}
\end{center}
\end{figure}

A heuristic to fill holes is to link each vertex lying on the hole to a
\emph{pseudo center} $\mathbf{c}$ of the hole.  This pseudo center corresponds
to the center of the circle associated to the hole, Fig. \ref{fig:filling}.  The
hole is approximated by a circle whose circumference $2\pi r$ corresponds to the
perimeter of the hole $\sum_j l_j$.  The vertices defining the hole are then
assumed to lie on such a circle.  New triangles are then defined, by connecting
those vertices to the pseudo center of the hole.  The angle $\alpha_j$ defined
by $\angle \mathbf{v_i} \mathbf{c} \mathbf{v_{i+1}}$ is assumed to be equal to
$\frac{l_j}{r}$.  Since the triangles filling the hole are assumed to share
$\mathbf{c}$, they are isosceles.  All those assumptions enable to average the
parametric coordinates of vertices lying on the hole, such that there was no
hole.  The triangles filling the hole are not explicitly built.

The heuristic performs well, even if the hole is concave and badly shaped,
Fig. \ref{sub:concaveFilling}.  The improvement compared to the homogeneous
Neumann condition is obvious, Fig. \ref{sub:concaveNeumann}.  Actually, some
parametric triangles of Fig. \ref{sub:concaveNeumann} are too tight for meshing
purposes.

The drawback of filling holes is that it increases the connectivity of the
linear system enabling the computation of the underlined parametrization.
Indeed, the pseudo centers correspond to extra unknowns which are related to
the corresponding unknowns along each hole.  Hence, the corresponding rows
within the matrix representing the linear system may have a lot of nonzero.
The corresponding linear system may become difficult to solve due to those latter rows.
In order to conserve a quick process of parameterization, a threshold of the potential connectivity is set: if there are too many vertices on a hole, homogeneous Neumann boundary conditions are set.
Otherwise, this hole is filled with the pseudo center.

\FloatBarrier
\section{Gmsh's Pipeline for Discrete Surface Meshing}
\label{sec:pipeline}

The specifications of Gmsh's algorithm for the generation of meshes on
discrete surfaces are  the following
\begin{itemize}
\item A conforming ``watertight'' geometrical triangulation is given as
  input.
\item A mesh with user specified mesh size parameters is given as
  output by Gmsh where all mesh vertices lie exactly on the input triangulation.
\end{itemize}

In Gmsh's new pipeline, the problem of surface meshing is divided in two stages:
(i) a pre-processing stage and (ii) a mesh generation stage.  In order to
explain the usefulness of the two stages of the pipeline, a relatively simple
example will serve as a common theme to illustrate the various treatements that
have to be undergone by a rough geometric triangulation to become a high quality
finite element mesh.

\begin{figure}[!ht]
\begin{center}
\includegraphics[width=.9\linewidth]{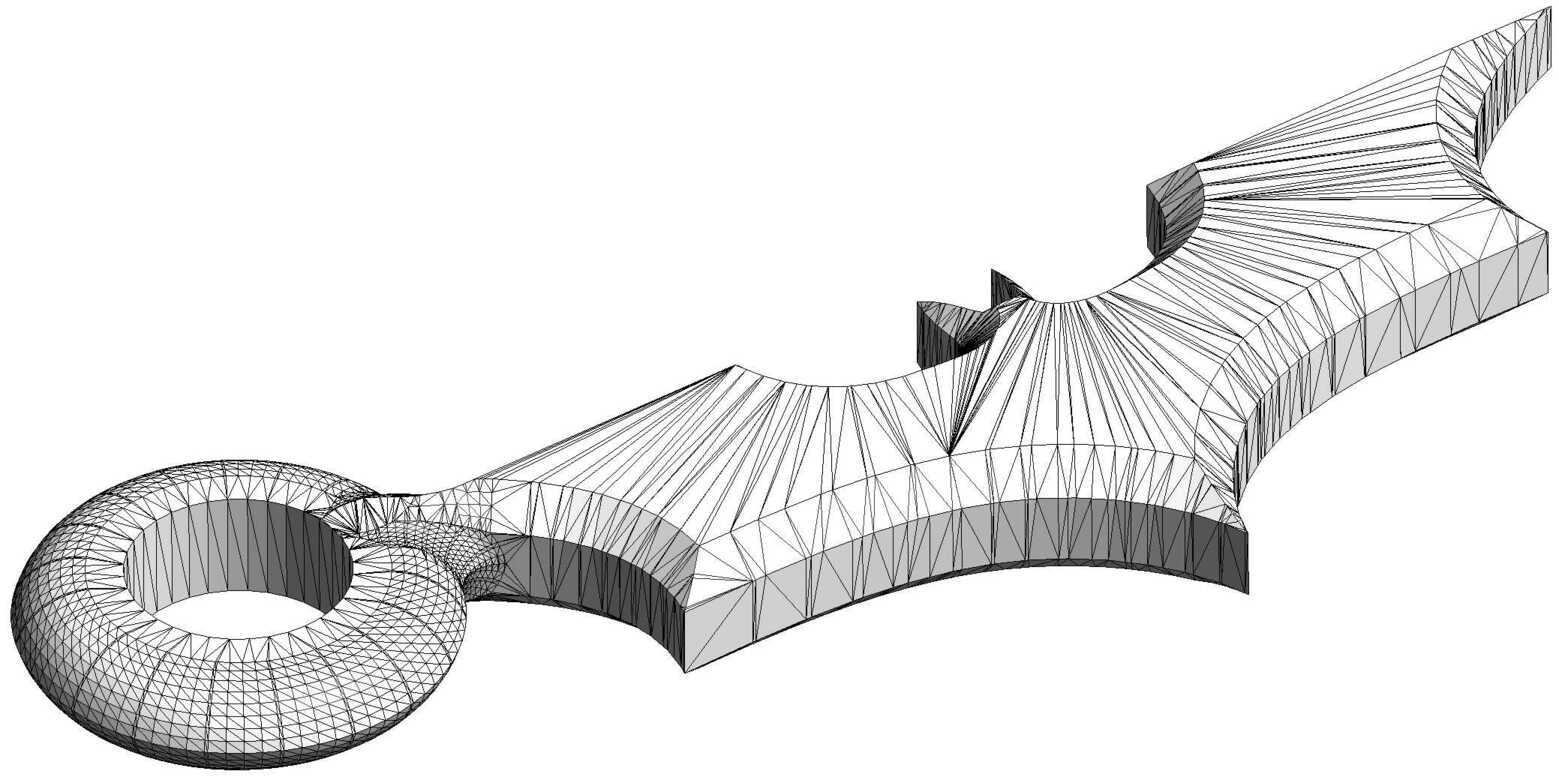}
\caption{The Batman geometry.}
\label{fig:batman_stl}
\end{center}
\end{figure}

Fig. \ref{fig:batman_stl} shows the geometric triangulation of a ``Batman''
object that is connected to a sort ot torus.

\FloatBarrier

In Gmsh's pipeling, a rough geometrical triangulation is taken as input. A
triangulation like the one of Fig. \ref{fig:batman_stl} cannot be processed as
is for a number of reasons.

\subsection{Detecting feature edges}

\begin{figure}[h!]
\begin{center}
\begin{tabular}{cc}
\includegraphics[width=.45\linewidth]{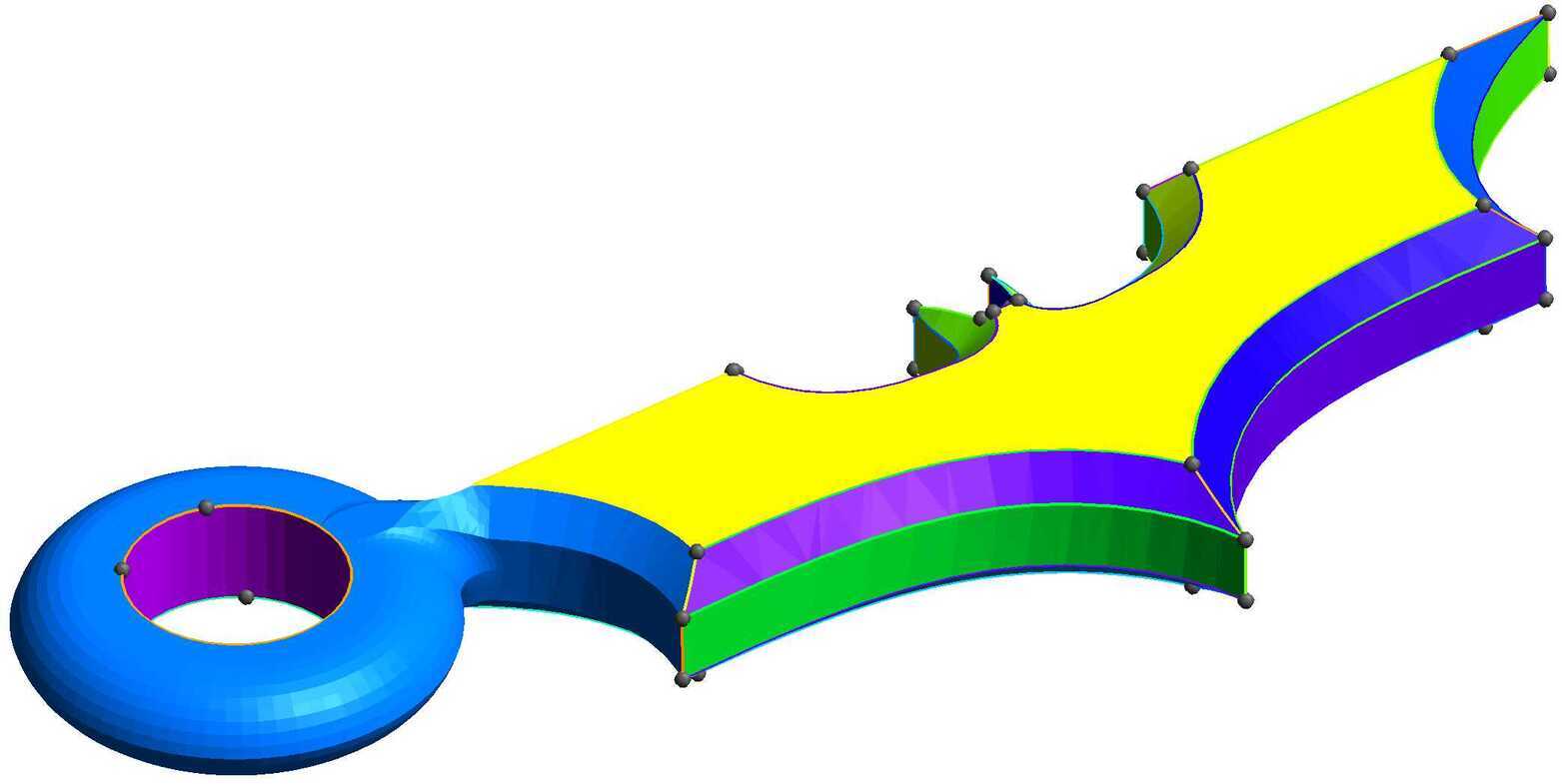}&
\includegraphics[width=.45\linewidth]{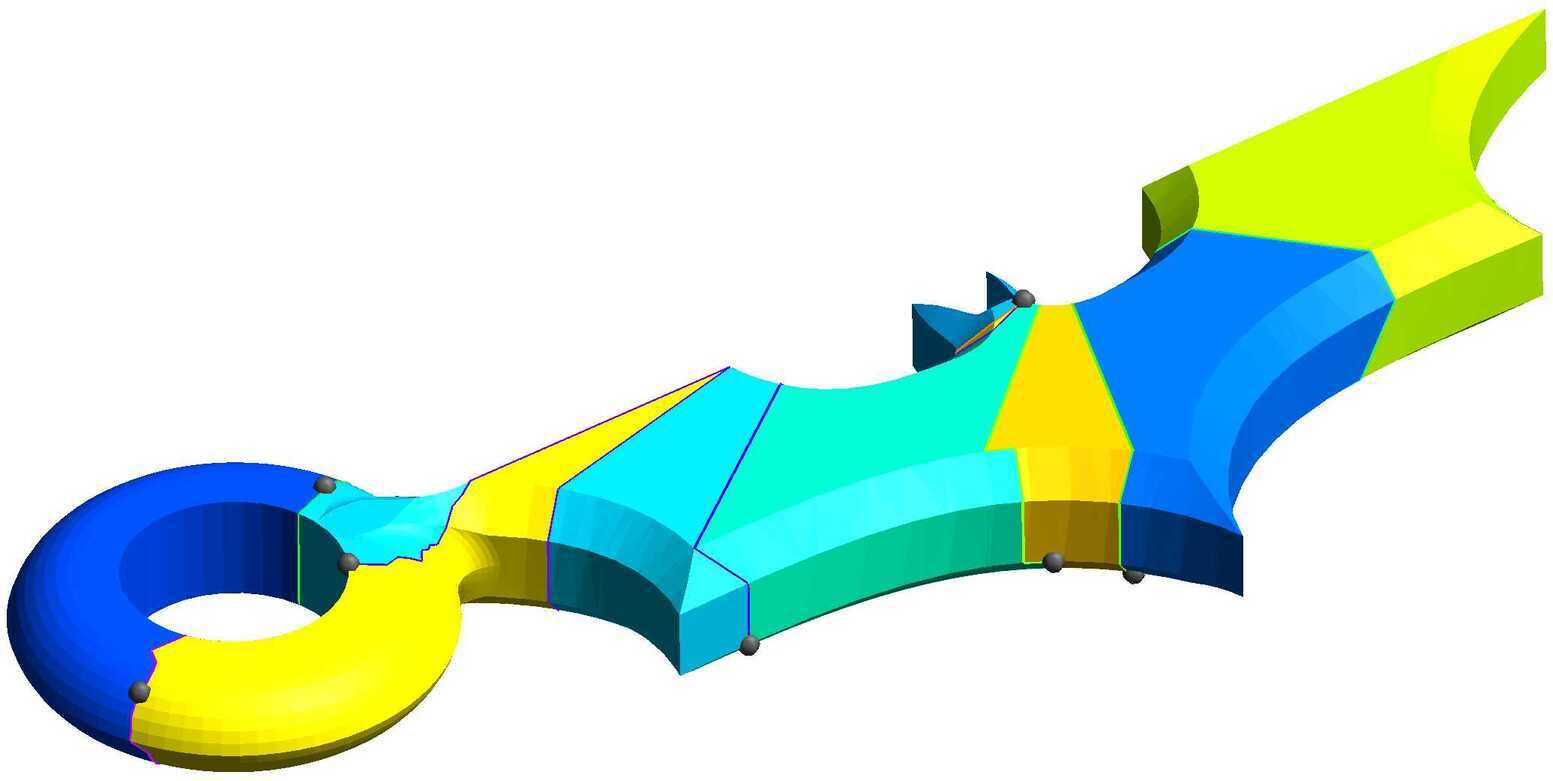}\\
\includegraphics[width=.45\linewidth]{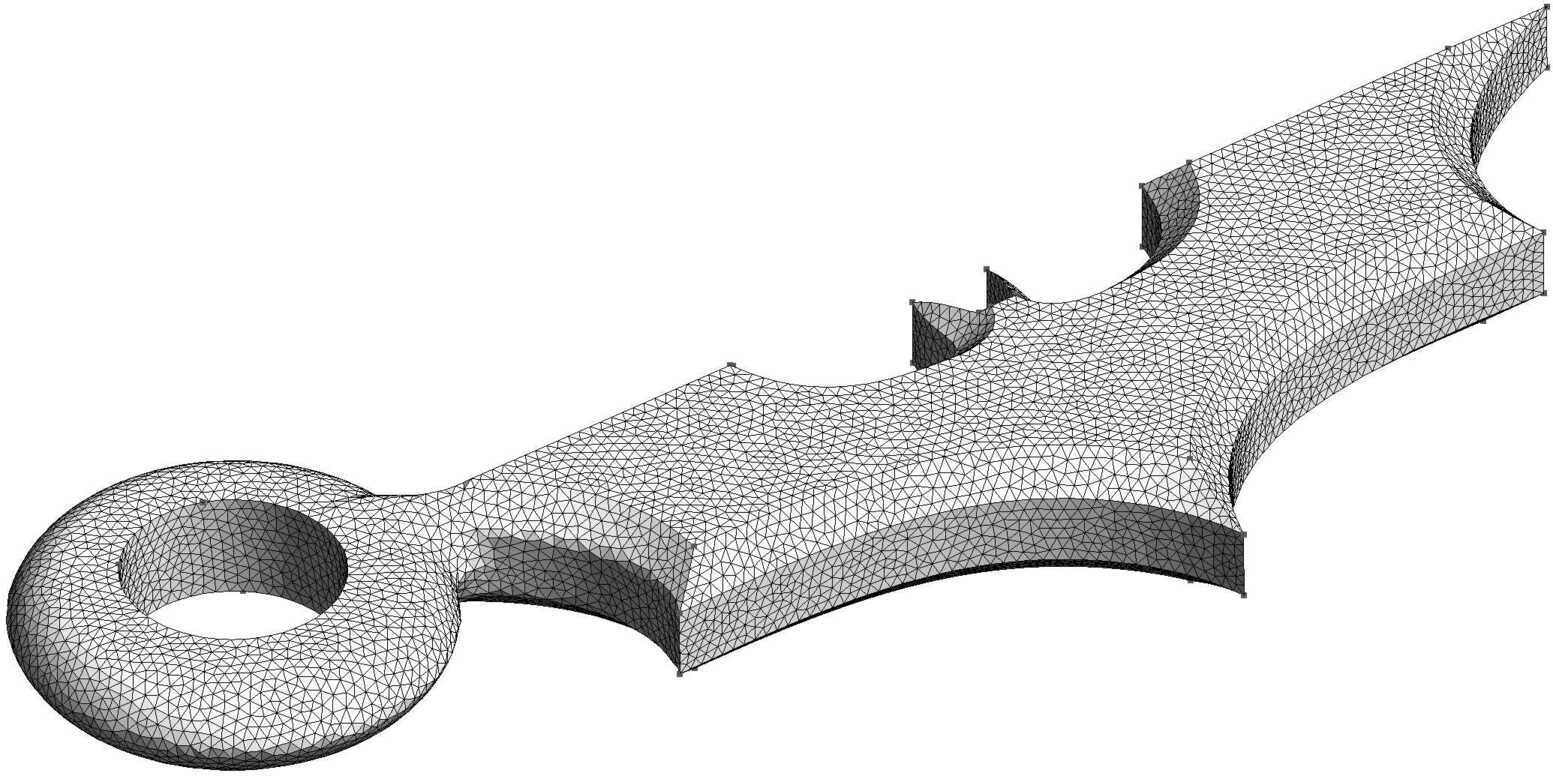}&
\includegraphics[width=.45\linewidth]{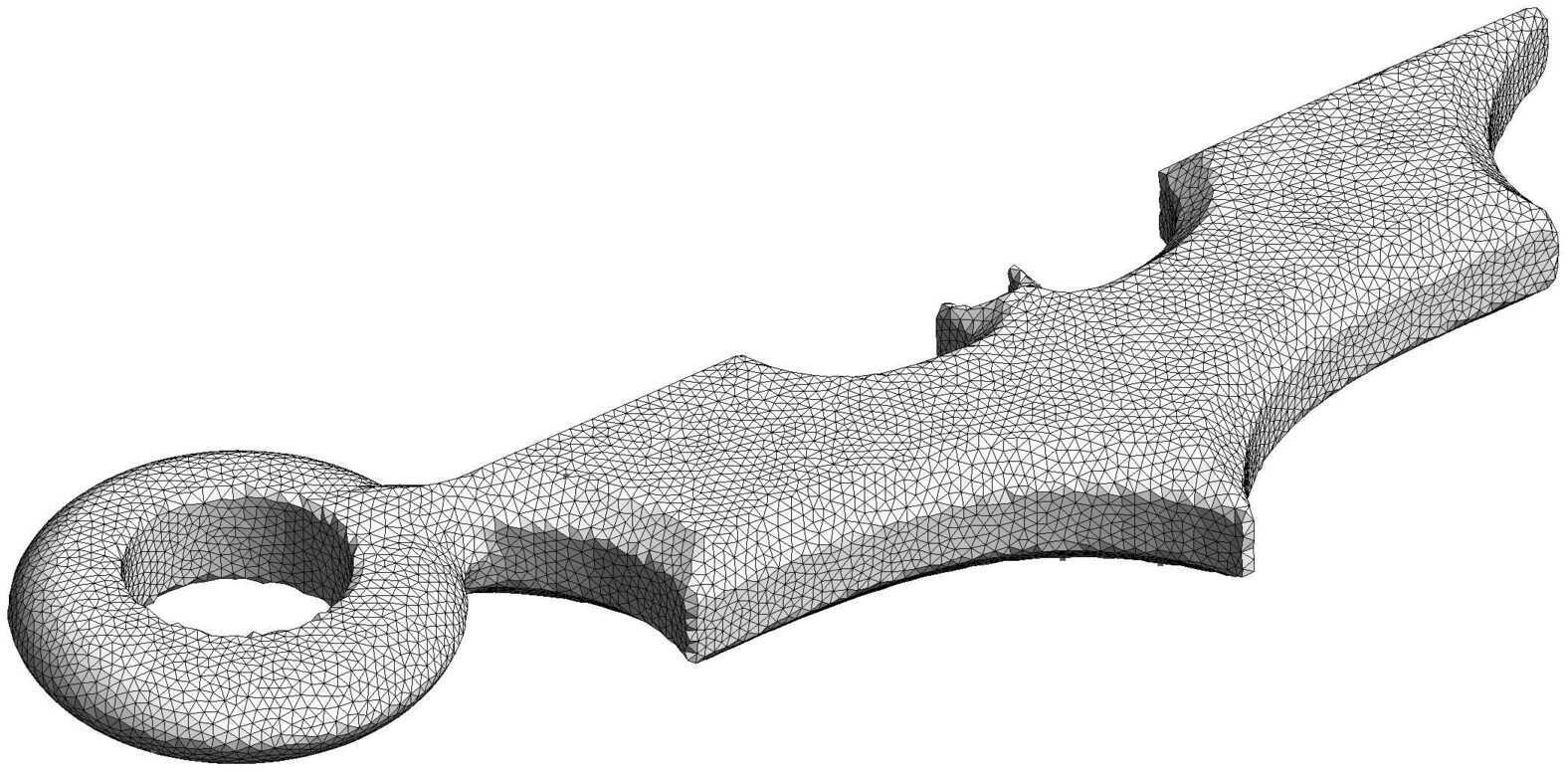}
\end{tabular}
\caption{Top left Fig. shows the final model with feature edges detection
  (threshold angle of 40 degrees). Bottom left Fig. shows a uniform mesh on
  that model. Right Fig. show the final model and mesh without feature edges
  detection. The domain has been split automatically in such a way that every
  model face has the right topology.}
\label{fig:batman_stage11}
\end{center}
\end{figure}

The geometrical triangulation of Fig. \ref{fig:batman_stl} is composed of a
list of triangles. The first part of our pre-processing is to detect
feature edges of the geometry that should be present in the final mesh. We use
here a simple angle criterion (typically, user-defined) to detect feature edges. After detecting feature
edges, a first version of the final atlas is created. Fig.
\ref{fig:batman_stage11} shows the Batman geometry where feature edges have been
created for all edges that have two adjacent triangles with normals separated by
an angle of more than $40$ degrees. A first version of the final topology of the
domain is created with model faces that are bounded by the feature edges.  After
the computation of feature edges, curvature tensors are computed at every vertex
of every surface using the method of \cite{rusinkiewicz2004estimating}.

\subsection{Creating the atlas}

At that point, we are not yet ready to compute the atlas of the model i.e. the
final boundary representation of our model together with the parametrization of
all its model surfaces. As explained in \S\ref{sec:discreteparam} every model
surface of the atlas should have the right topology. In this following step, we
ensure that every map of the atlas has this right topology. When a surface has a
larger genus, it is split in two parts using \texttt{METIS} (see \cite{METIS}), a graph partitioning software.
This operation is applied up to the point when every surface is parametrizable.

It is also known that surfaces with large aspect ratios may lead to
parametrizations that have non distinguishable coordinates.  When the
parametrization is computed, we also ensure that parametric triangles are not
too small i.e that their area is not close to machine precision
(see \cite{marchandise2011high}).  If it is the case, the surface is split in two.

For large models, we also split surfaces that contain a too large number of
triangles (typically 100,000). Computing mean value coordinates require to solve
a non symmetrical system of equations and one of the design goals of the
parametrization process is to be fast.

Fig. \ref{fig:batman_stage11} (top right) shows the decomposition that has
been done on the Batman model without pre-computing feature edges.

\subsubsection{The final BREP}

At that point, the input triangulation has been transformed into a proper
boundary representation that has a valid topology and for which each face has
been parametrized. All those topological and geometrical informations are now
saved in the version 4 of the output mesh format of Gmsh. This ``extended'' mesh
file can be used as input to Gmsh's surface mesh generators. Fig.
\ref{fig:batman_stage11} (bottom images) show meshes for both models generated
using feature edges and automatic splitting.

\FloatBarrier
\section{Improving Parametrization on Coarse Discrete Surfaces}
\label{sec:coarse}

\begin{figure}[ht!]
\begin{center}
\subfloat[Parametrization ($u$: red isolines, $v$: blue isolines).]{\includegraphics[width = .45\linewidth]{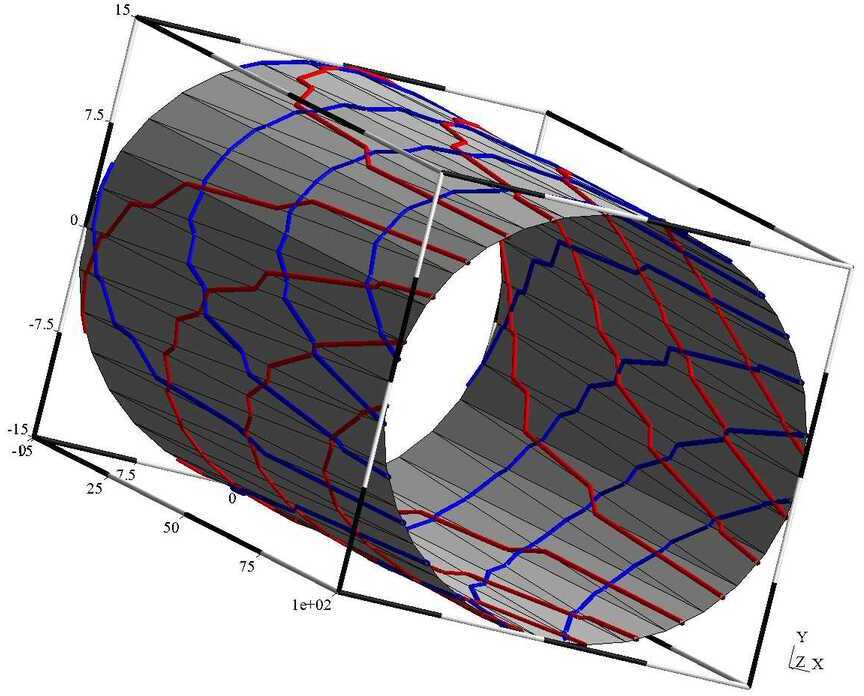}\label{sub:cylinder}}\hspace*{.25cm}
\subfloat[Color map: $\dfrac{\sigma_2}{\sigma_1}$.]{\includegraphics[width = .45\linewidth]{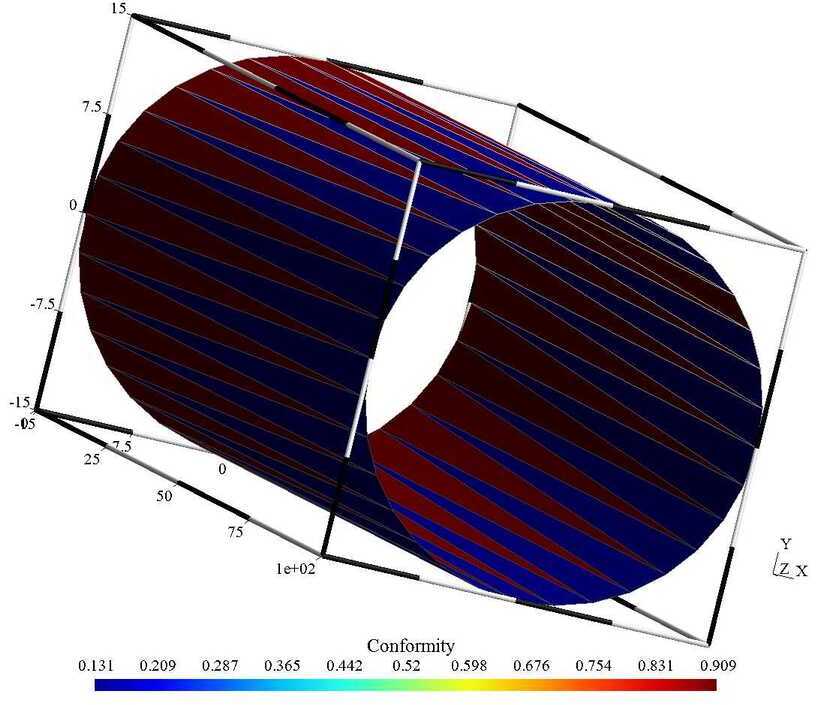}\label{sub:conformality}}\\
\subfloat[Parametric space.]{\includegraphics[width = .45\linewidth]{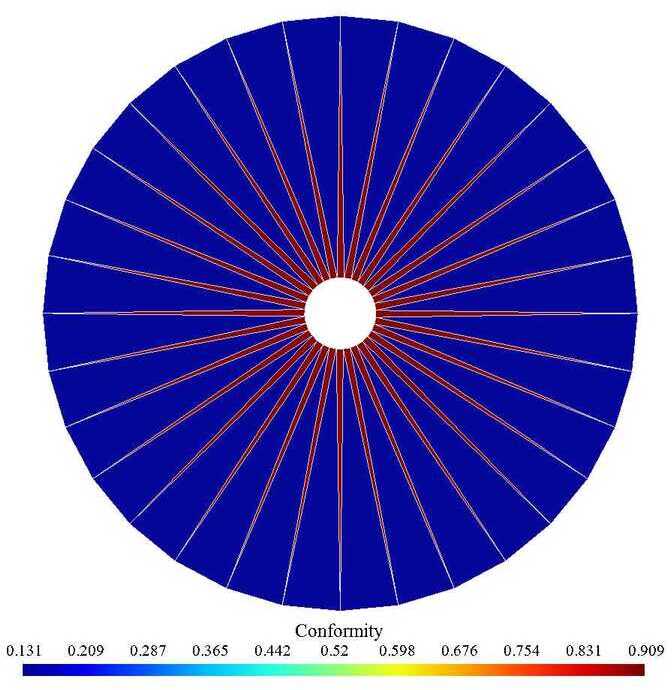}\label{sub:cylinderDisk}}
\end{center}
\caption{Parametrization on coarse stl triangulations: a cylinder.}
\label{fig:stl}
\end{figure}

The methodology that has been presented before is general and applies to
triangulated surfaces of arbitrary complexity. Yet, geometrical triangulations
of CAD surfaces may not be sufficiently dense to allow a smooth
parametrization. For example, a good geometrical triangulation of a cylinder may
not contain internal vertices as depicted on Fig. \ref{fig:stl}. We
have computed the parametrization of this cylinder using mean value coordinates
and the result is presented in Fig. \ref{sub:cylinder}.  Even if this
parametrization is said ``moderately noised'', it cannot be used for mesh
generation purposes. Fig. \ref{sub:conformality} and \ref{sub:cylinderDisk}
show conformity indicator $\frac{\sigma_2}{\sigma_1}$ both on the real and
parameter space of the cylinder.

From this observation, a numerical analyst would suggest two ways to improve the
computation: refining the solution (i.e. the input mesh), or increasing the
order of the approximation (i.e. second order).

\FloatBarrier
\subsection{Refinement by longest edge bisection}

\begin{figure}[ht!]
\begin{center}
\subfloat[Parametrization ($u$: red isolines, $v$: blue isolines).]{\includegraphics[width = .45\linewidth]{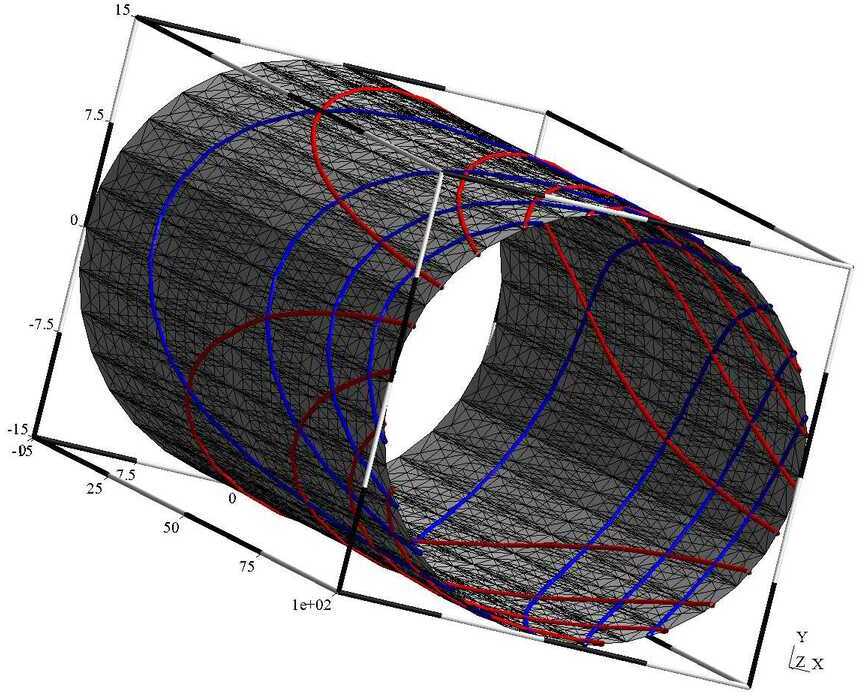}\label{sub:rcylinder}}\hspace*{.25cm}
\subfloat[Color map: $\dfrac{\sigma_2}{\sigma_1}$.]{\includegraphics[width = .45\linewidth]{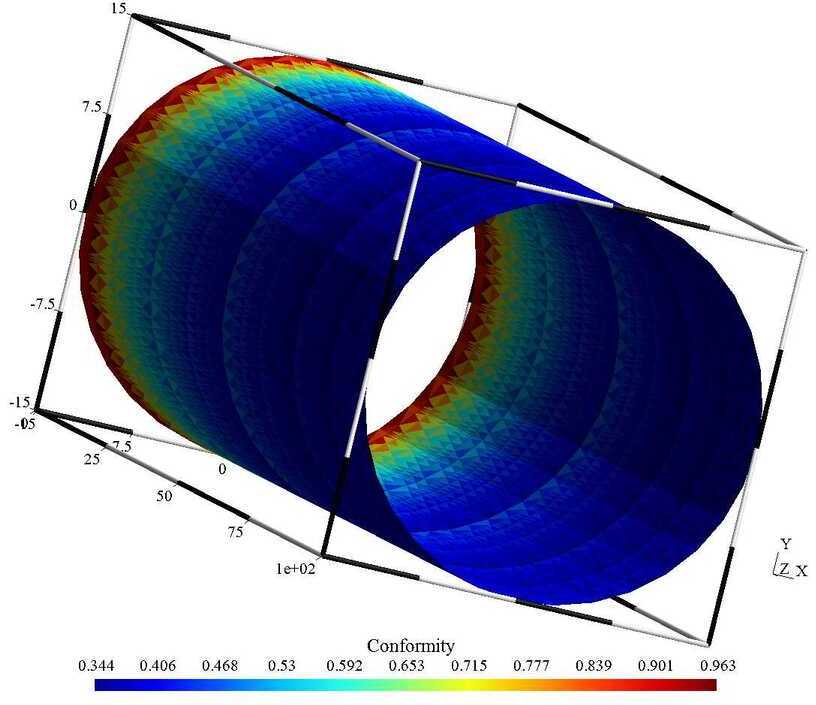}\label{sub:rconformality}}\\
\subfloat[Parametric space.]{\includegraphics[width = .45\linewidth]{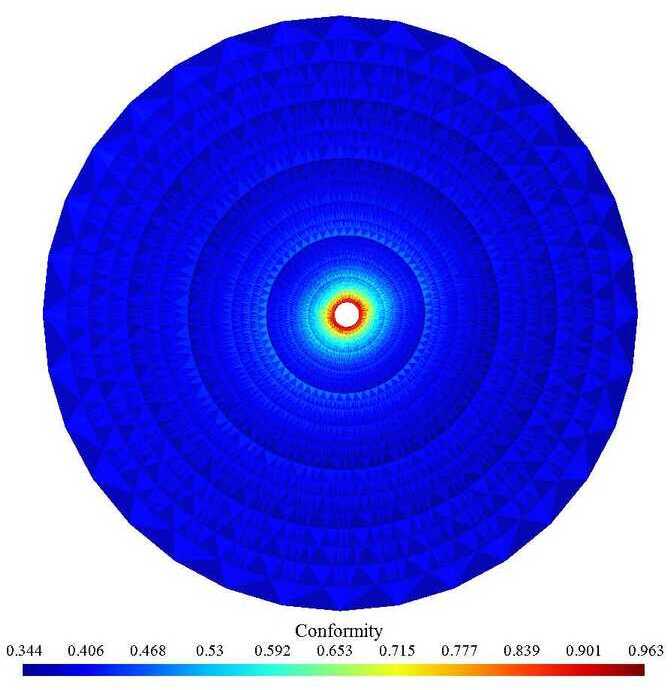}\label{sub:rcylinderDisk}}
\end{center}
\caption{Parametrization on refined stl triangulations: the cylinder (5 iterations).}
\label{fig:rstl}
\end{figure}

We refine the geometrical triangulation without changing its geometry i.e. only
using edge splits.  We use here a variant of the well known longest edge
bisection process developed by \citep{rivara1997new}: edges to be split are tagged and the
longest edge of the list is split, then the second longest edge is split and the
process continues until the shortest edge of the list is split.  We repeat the
process several times up to the point all inner edges respect a length
threshold. Fig. \ref{sub:rcylinder} shows the new geometrical mesh of the
cylinder.

\begin{figure}[!ht]
\begin{center}
\subfloat[Good mesh on cylinder.]{
\begin{tikzpicture}[every node/.style={inner sep=0,outer sep=0}]
\node at (0,0) {\includegraphics[width = .45\linewidth]{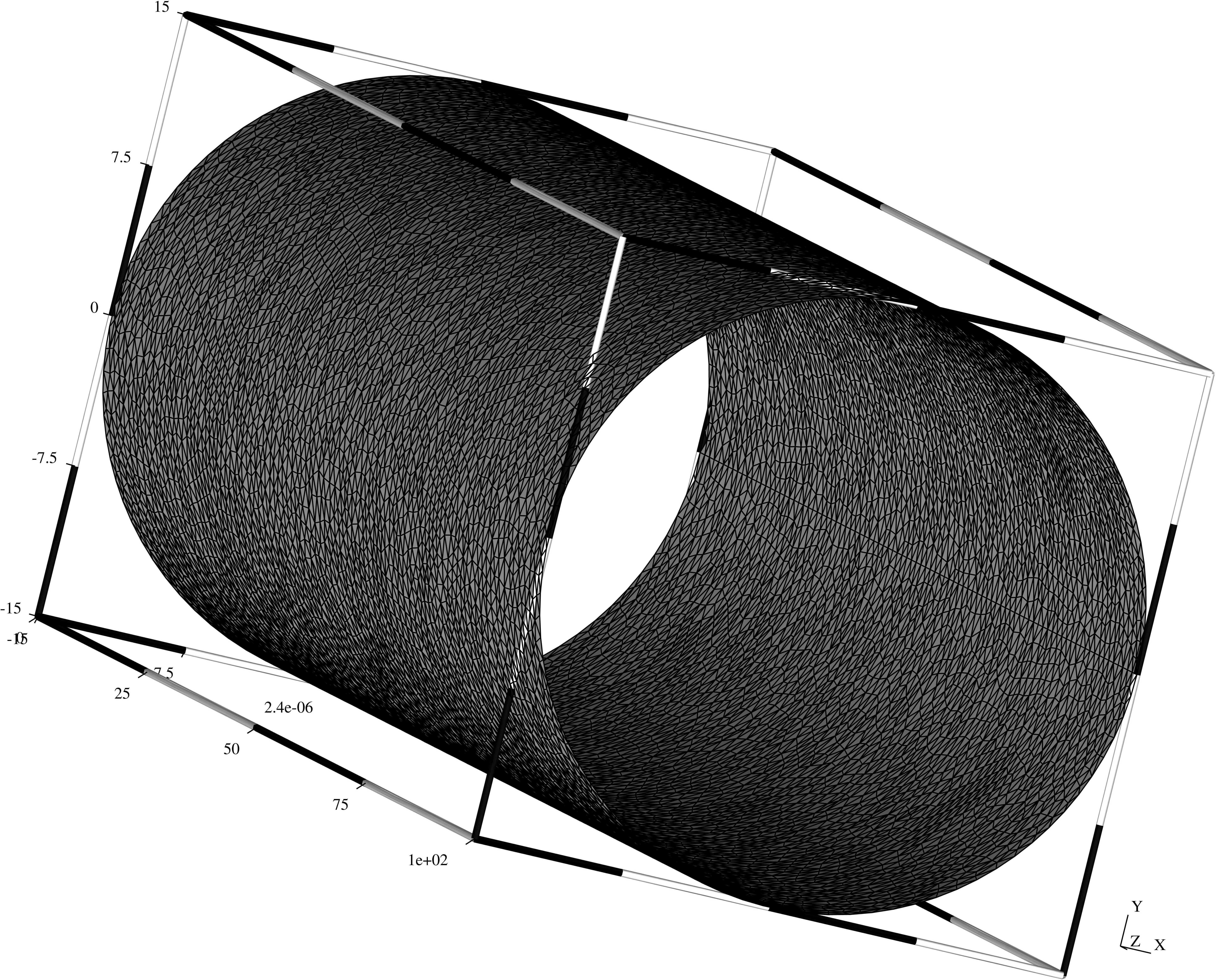}\label{sub:fineCylinder}};
\draw[very thick,ForestGreen] (-.9,-0.4)--(-1.7,-0.4)--(-1.7,0.15)--(-.9,0.15)--cycle;
\node[draw, line width=1mm,ForestGreen] at (-6,0) {\includegraphics[width = .45\linewidth]{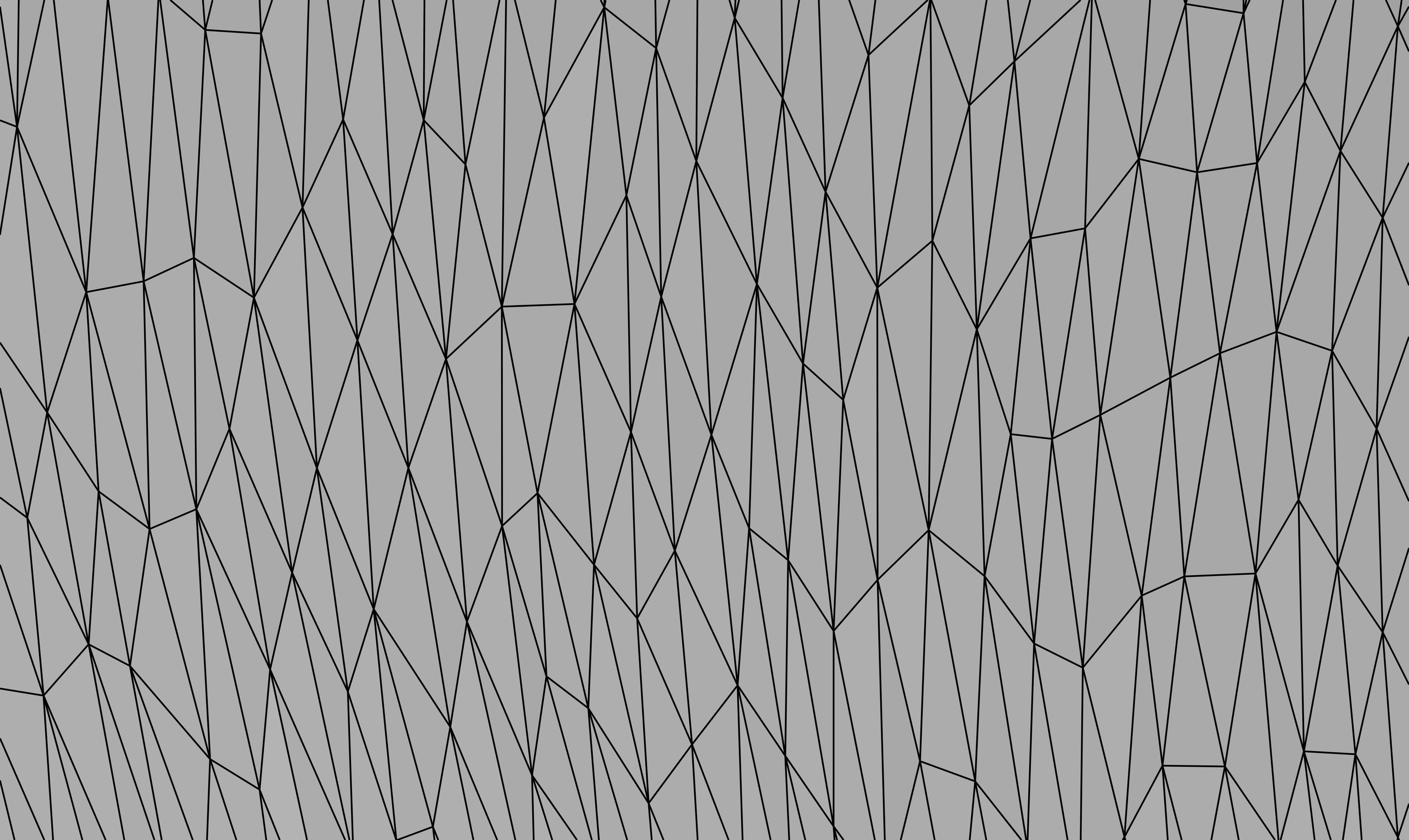}};
\end{tikzpicture}
}\\
\subfloat[Mapping within the parametric space, \emph{without} preprocessing (cf. Fig. \ref{sub:cylinderDisk}).]{
\begin{tikzpicture}[every node/.style={inner sep=0,outer sep=0}]
\node at (0,0) {\includegraphics[width = .45\linewidth]{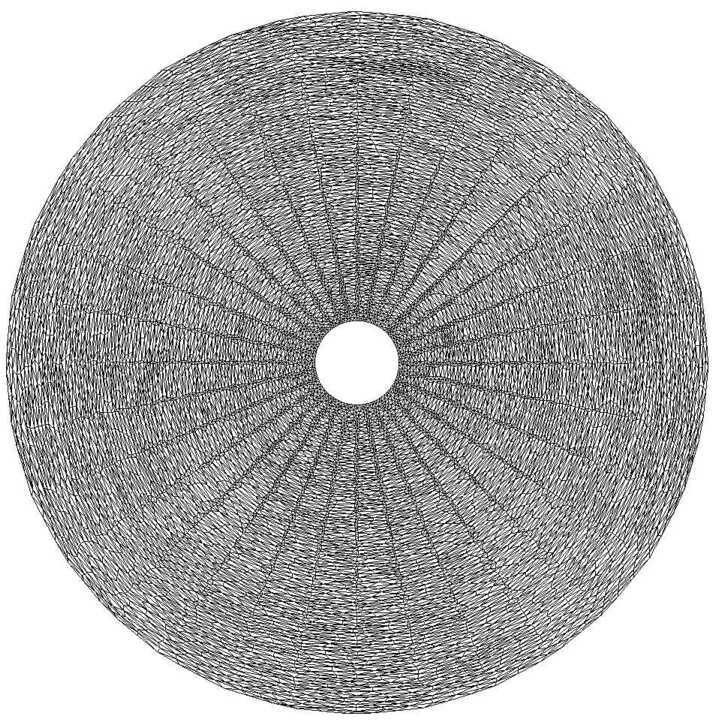}\label{sub:badMapped}};
\draw[very thick,ForestGreen] (.1,.1)--(2.5,.1)--(2.5,1.6)--(.1,1.6)--cycle;
\node[draw, line width=1mm,ForestGreen] at (6,0) {\includegraphics[width = .45\linewidth]{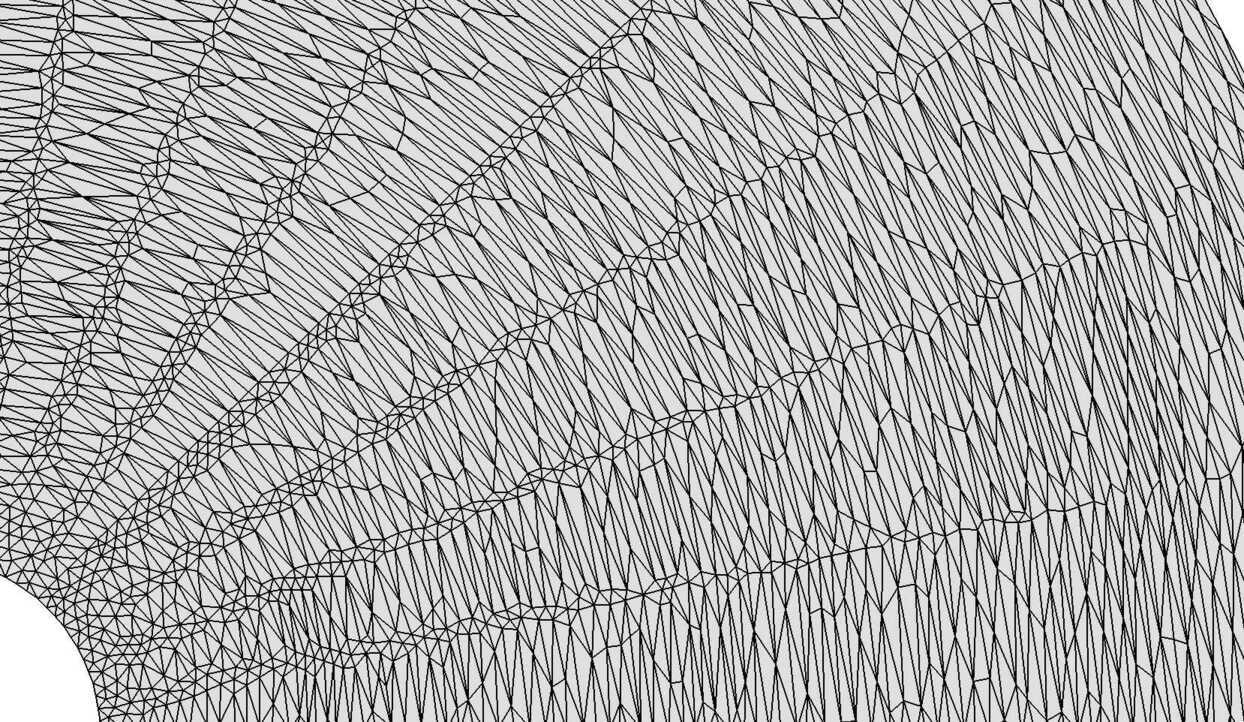}};
\end{tikzpicture}
}\\
\subfloat[Mapping within the parametric space, \emph{with} preprocessing (cf. Fig. \ref{sub:rcylinderDisk}).]{
\begin{tikzpicture}[every node/.style={inner sep=0,outer sep=0}]
\node at (0,0) {\includegraphics[width = .45\linewidth]{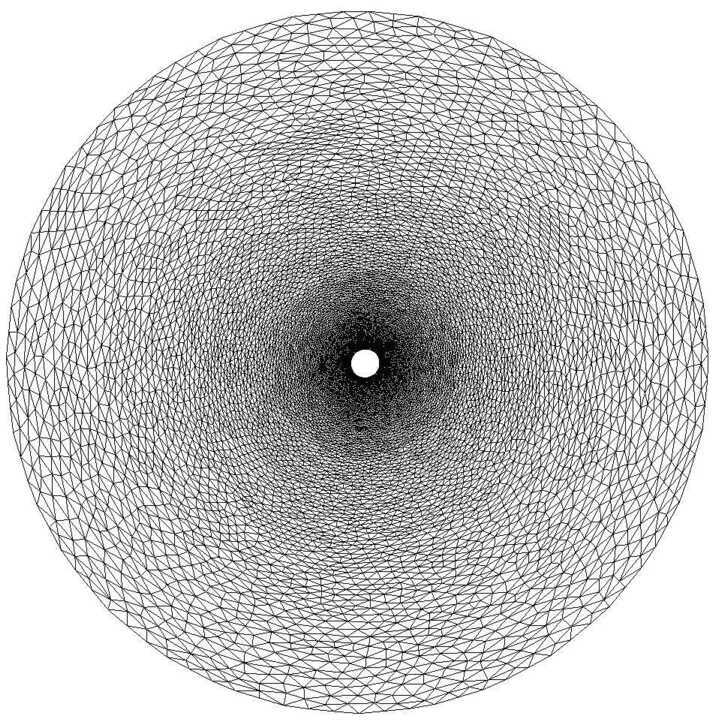}\label{sub:goodMapped}};
\draw[very thick,ForestGreen] (.1,0)--(2.5,0)--(2.5,1.6)--(.1,1.6)--cycle;
\node[draw, line width=1mm,ForestGreen] at (6,0) {\includegraphics[width = .45\linewidth]{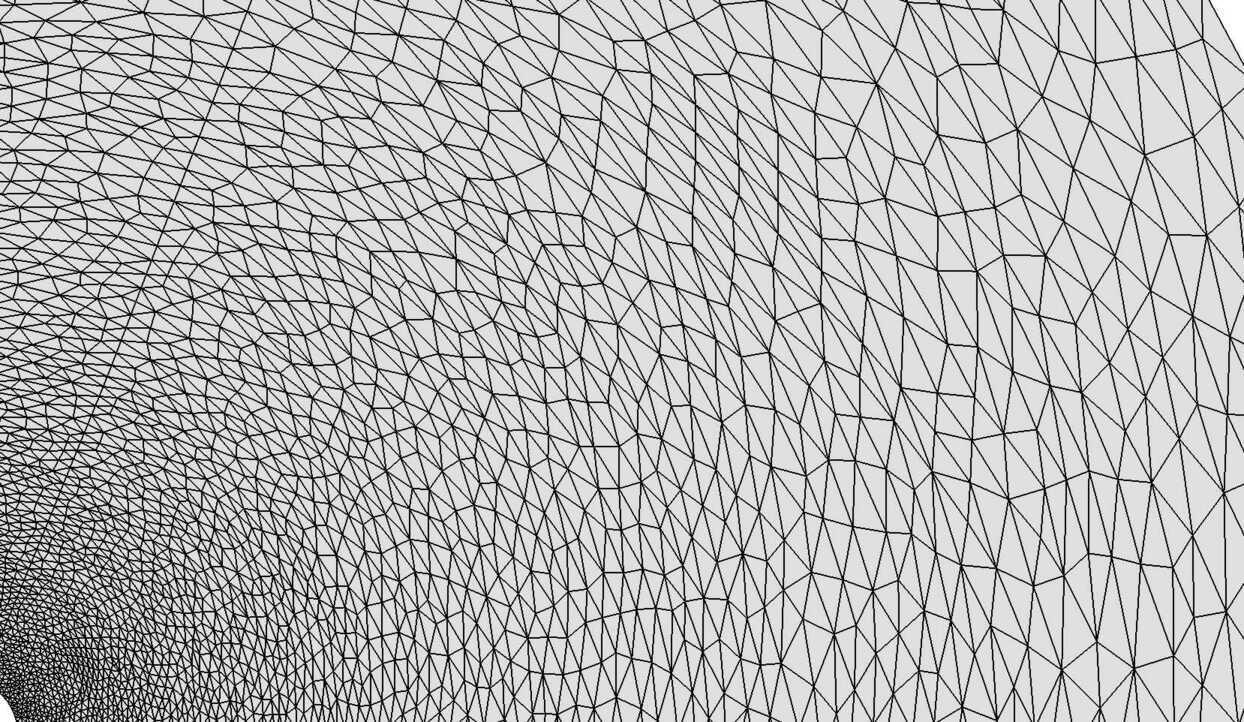}};
\end{tikzpicture}
}
\end{center}
\caption{Effect on mapping a good mesh on a parametrization with (b) and without (c) edge refinement preprocessing.}
\label{fig:mappingMesh}
\end{figure}

In order to illustrate the effect of this refinement on the
parametrization, we have pre-computed a ``good mesh'' of the cylinder
in the 3D space (see Fig. \ref{sub:fineCylinder}).
This good mesh has been inverse-mapped onto the parameter
spaces of the non refined cylinder and on the refined cylinder. While
doing that, we can see the meshes that should have been created by
Gmsh's surface meshers in both parameter planes to obtain the same ``good
mesh''. Fig. \ref{sub:badMapped} shows the mesh in the parameter
plane of the non refined geometrical cylinder: it contains series of
elongated triangles followed by isotropic ones, illustrating the too
great variability of the conformity parameter. In Fig.
\ref{sub:goodMapped}, the mesh is anisotropic but element shapes are
locally uniform and any good anisotropic mesher is able to generate
that kind of mesh.

\FloatBarrier
\subsection{Second order approximation}

\begin{figure}[!ht]
\begin{center}
\includegraphics[width=.5\linewidth]{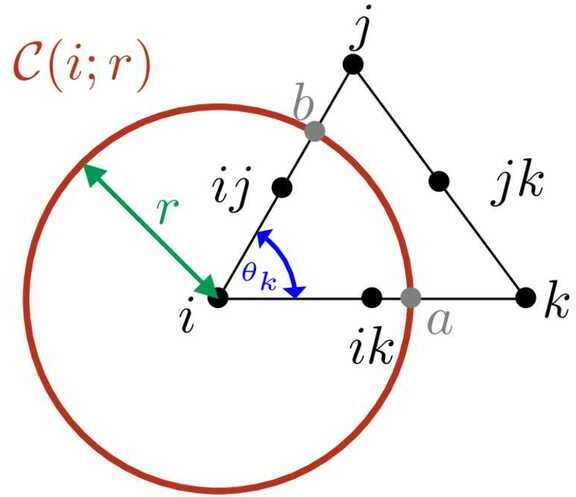}
\caption{Sketch for quadratic approximation of $\lambda^{\mathcal{P}^2}_{ij}$.}
\label{fig:lamdbda2}
\end{center}
\end{figure}

As in the piecewise linear approximation (see §\ref{sec:mvc}), we derive $\lambda_{ij}$ from Lagrange $\mathcal{P}^2$ function shapes
$$
\theta_k r ~ f_i = \int_{\arc{ab}} f_i \phi_i + f_j \phi_j +  f_k \phi_k + f_{ij} \phi_{ij} + f_{jk} \phi_{jk} + f_{ik} \phi_{ik} ~ ds
$$
where $\phi_{\bullet}$ are the Lagrange $\mathcal{P}^2$ finite element shape functions, which are defined with the barycentric coordinates $(\mathtt{v}_i,\mathtt{v}_j,\mathtt{v}_k)$ \citep[Chapter 1,§1.2.4]{ern2013theory}
$$
\left\{\begin{array}{rcl}
\phi_a &=& \mathtt{v}_a (2\mathtt{v}_a-1),~a\in\{i,j,k\}\\
\phi_{ab} &=& 4\mathtt{v}_a \mathtt{v}_b,~ a,b \in \{i,j,k\}:a \neq b
\end{array}\right.
$$
Assigning coordinates relative to $v_i$, Fig. \ref{fig:lamdbda2}
$$
\begin{array}{rcl}
\mathtt{v}_i &=& (0;0)\\
\mathtt{v}_j &=& (l_{ij}\cos(\theta_k);l_{ij}\sin(\theta_k))\\
\mathtt{v}_k &=& (l_{ik};0)\\
\end{array}
$$

Again, $\phi_i+\phi_j+\phi_k+\phi_{ij}+\phi_{jk}+\phi_{ik}=1$ enables us to write

\begin{align*}
\underbrace{\left(\theta_k r - \int_{\arc{ab}} \phi_i ~ ds \right)}_{\int_{\arc{ab}} \phi_j + \phi_k + \phi_{ij} + \phi_{jk} + \phi_{ik} }f_i - \int_{\arc{ab}} \phi_j ~ds f_j - \int_{\arc{ab}} \phi_k ~ds f_k \\ - \int_{\arc{ab}} \phi_{ij}~ds f_{ij} - \int_{\arc{ab}} \phi_{jk}~ds f_{jk} - \int_{\arc{ab}} \phi_{ik}~ds f_{ik} = 0
\end{align*}

which gives
\begin{align*}
\underbrace{\int_{\arc{ab}} \phi_j ~ds}_{\lambda_{ij}}(f_i-f_j) + \underbrace{\int_{\arc{ab}} \phi_k ~ds}_{\lambda_{ik}}(f_i-f_k) \\ + \underbrace{\int_{\arc{ab}} \phi_{ij} ~ds}_{\lambda_{i(ij)}}(f_i-f_{ij}) + \underbrace{\int_{\arc{ab}} \phi_{jk} ~ds}_{\lambda_{i(jk)}}(f_i-f_{jk}) + \underbrace{\int_{\arc{ab}} \phi_{ik} ~ds}_{\lambda_{i(ik)}}(f_i-f_{ik}) = 0
\end{align*}

We use \texttt{SymPy} (see \citep{meurer2017sympy}) to compute $\lambda^{\mathcal{P}^2}_{ij}$ (code in supplementary material)
\begin{equation}\label{eqn:lambda2}
\lambda^{\mathcal{P}^2}_{ij} = \dfrac{r^2}{l_{ij}^2 ~ \sin^2(\theta_k)} \left( (l_{ij} -r ) \cos(\theta_k) \sin(\theta_k) + r \theta_k - l_{ij} \sin(\theta_k)\right)
\end{equation}

We should derive the other coefficients $\lambda^{\mathcal{P}^2}_{\bullet}$, but something is wrong with \eqref{eqn:lambda2}.
We cannot get rid of $r$ within the expression.
It means that the coefficients give the average for a certain circle of radius $r$.
Yet, it has to be for \emph{any} circle, whatever the radius.
It is then not possible to derive $\lambda^{\mathcal{P}^2}$ for a monotone scheme.

\begin{figure}[!ht]
\begin{center}
\subfloat[Kuratowski graph of type \texttt{I}.]{\includegraphics[width=.3\linewidth]{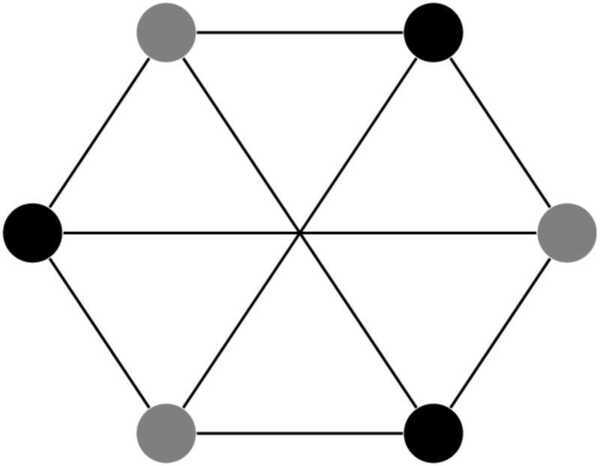}\label{sub:kuratowski}}\hspace*{.25cm}
\subfloat[Graph depicting Lagrange $\mathcal{P}^2$ dof's on a triangle.]{\includegraphics[width=.3\linewidth]{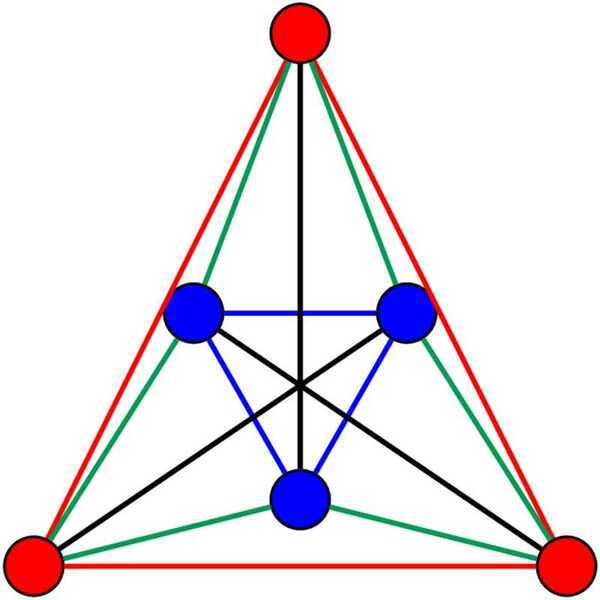}\label{sub:lagrangep2tri}}\hspace*{.25cm}
\subfloat[Kururatowski subgraph of Lagrange $\mathcal{P}^2$ dof's on a triangle.]{\includegraphics[width=.3\linewidth]{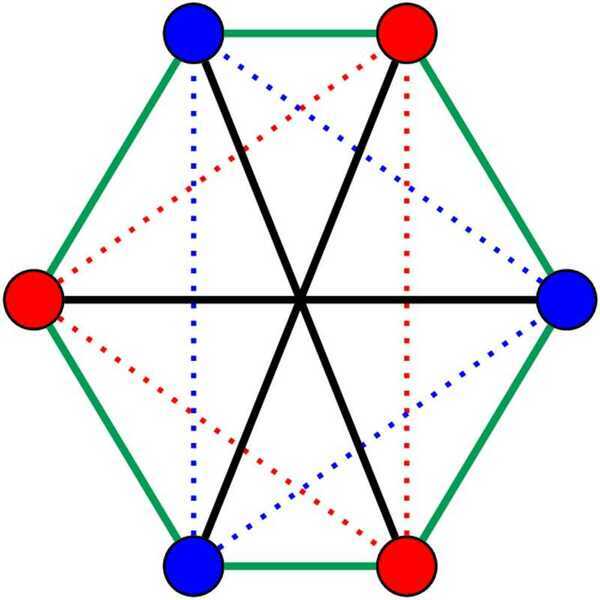}\label{sub:kuratowskip2tri}}\\
\end{center}
\caption{Graph corresponding to Lagrange $\mathcal{P}^2$ dof's on a triangle has no planar representation.}
\label{fig:kuratowskip2}
\end{figure}

Actually, graph theory states such a result.  Lagrange $\mathcal{P}^2$ degrees
of freedom on a triangle may be depicted by a 3-connected graph,
Fig. \ref{sub:lagrangep2tri}.  Tutte \citep[§4]{tutte1963draw} claims that any
graph having a Kuratowski subgraph is nonplanar.  Fig. \ref{sub:kuratowski}
corresponds to a Kuratowski graph.  A graph is planar if it can be drawn on a
plane, in such a way that its edges intersect only on vertices of the graph.  It
means that each vertex of the graph may correspond to a convex combination of
its neighbors, which we aim.  However, Fig. \ref{sub:lagrangep2tri} has such a
Kuratowski subgraph, Fig. \ref{sub:kuratowskip2tri}.  The graph of
Fig. \ref{sub:lagrangep2tri} has no planar representation.  Hence, it means it
is not possible to write Lagrange $\mathcal{P}^2$ scheme which is monotone.

\FloatBarrier
\section{Examples}
\label{sec:examples}

In this section, several complex examples are presented that show the
level of robustness that has been attained by our methodology. The
examples that have been chosen in order to challenge our algorithm and
push it to the limit.

\begin{figure}[!ht]
\begin{tabular}{cc}
\includegraphics[width=.45\linewidth]{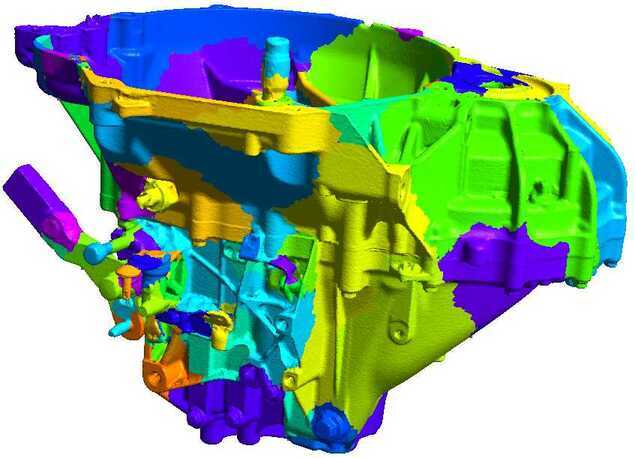} &  \includegraphics[width=.45\linewidth]{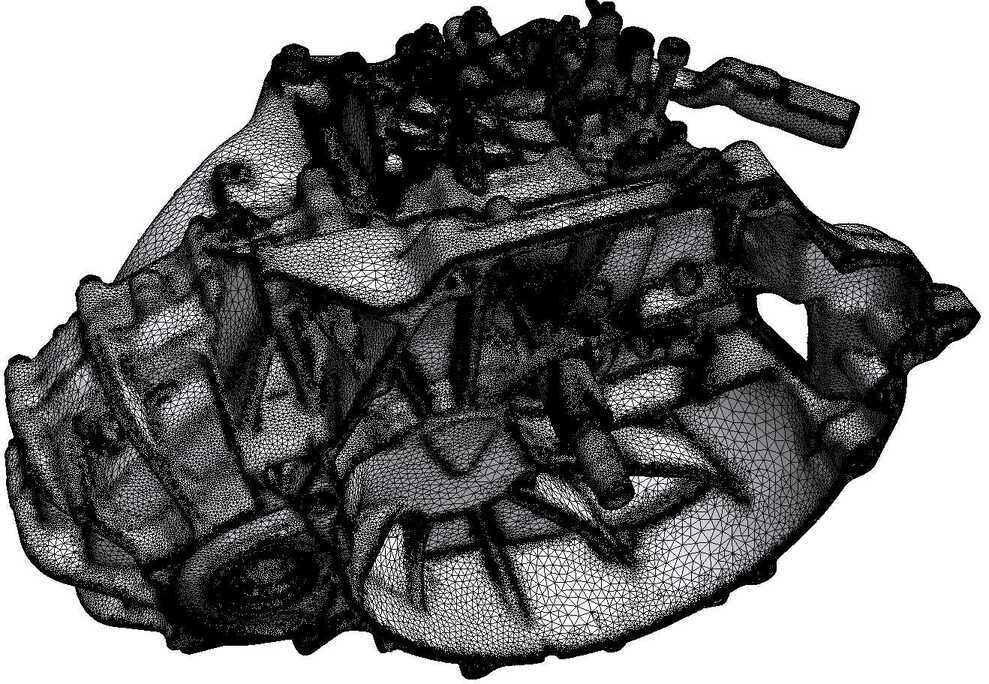}
\end{tabular}
\caption{Complex scanned mechanical part. The initial triangulation (left) that contains
  $797,666$ triangles has been split into $194$ surfaces that are
  parametrizable. The  mesh on the right that contains $1,762,388$
  triangles and has been adapted to the curvature of the original
  discrete surface. It has been generated by Gmsh in $640$ seconds, including
  IO's. \label{fig:ARK}}
\end{figure}

\begin{figure}[!ht]
\begin{tabular}{cc}
\includegraphics[width=.45\linewidth]{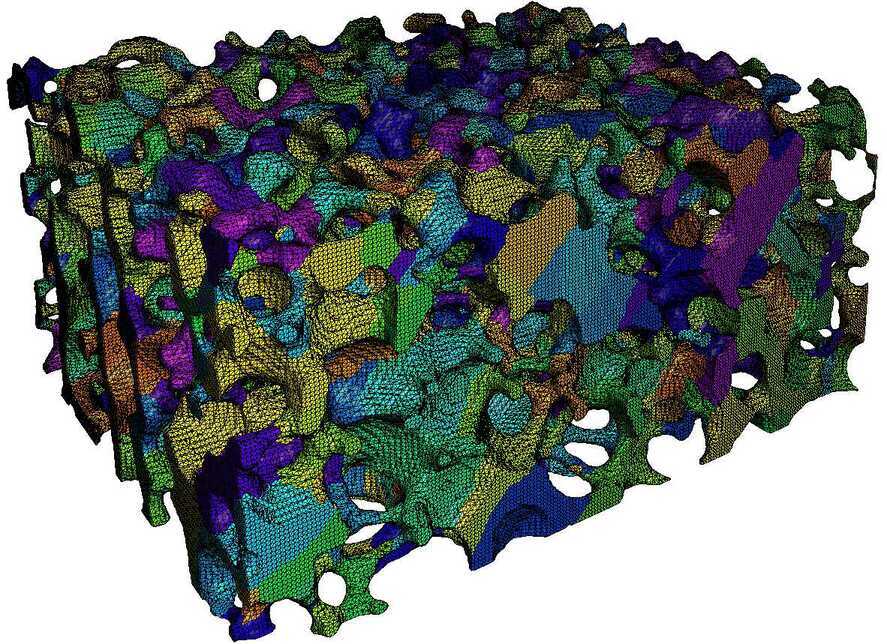} &  \includegraphics[width=.45\linewidth]{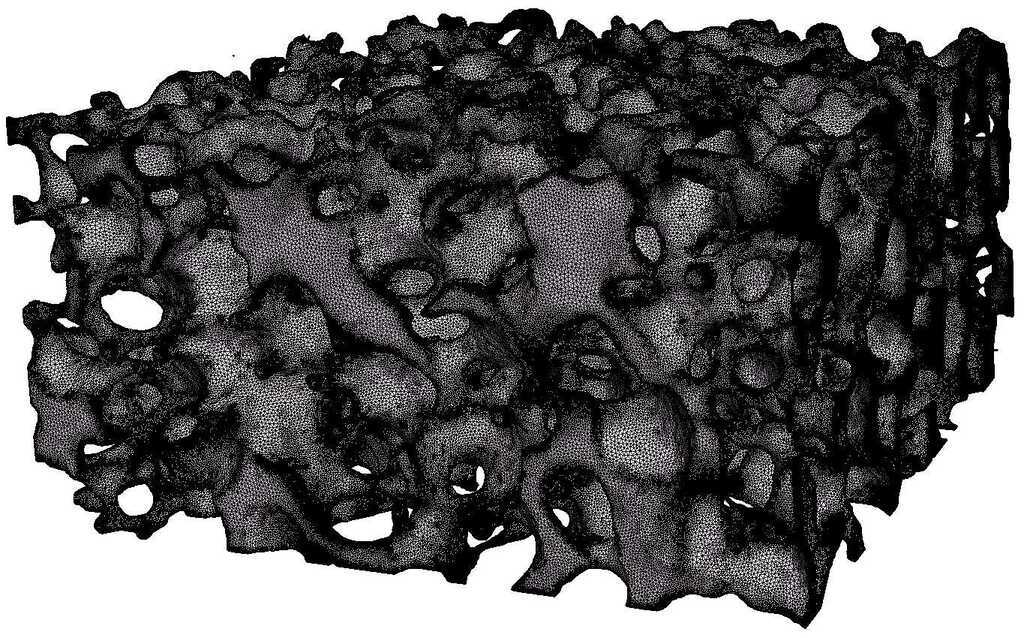}
\end{tabular}
\caption{X-ray tomography image of a silicon carbide foam (from P. Duru,
  F. Muller and L. Selle, IMFT, ERC Advanced Grant SCIROCCO). The initial
  triangulation (left) that contains $1,288,116$ triangles has been split into
  $1,802$ surfaces that are parametrizable. The mesh on the right contains
  $4,922,322$ triangles and has been adapted to the curvature of the original
  discrete surface. It has been generated by Gmsh in $1,187$ seconds, including
  IO's. \label{fig:morph}}
\end{figure}
\begin{figure}[!ht]
\begin{tabular}{cc}
\includegraphics[width=.45\linewidth]{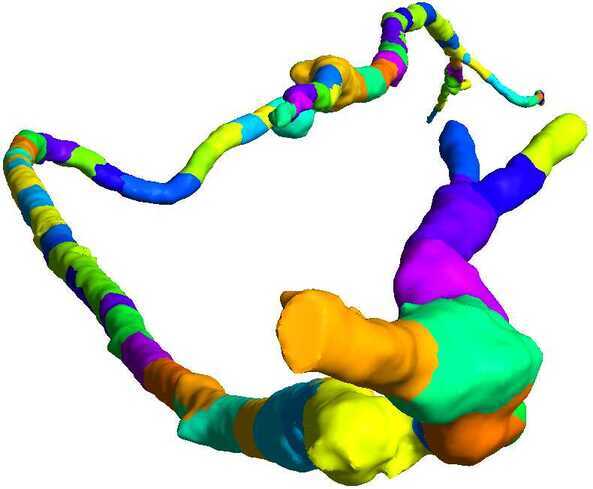} &  \includegraphics[width=.45\linewidth]{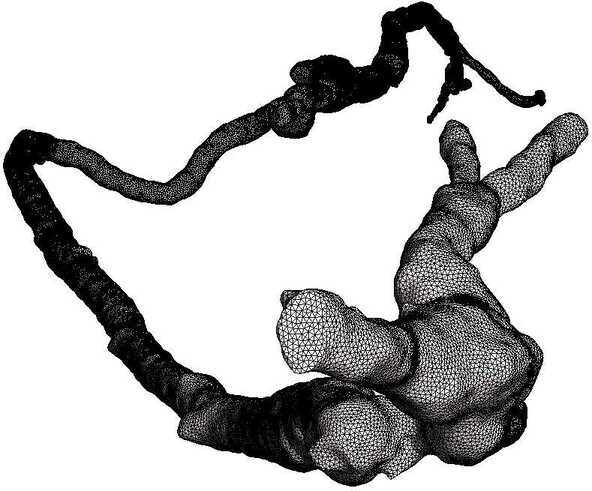}
\end{tabular}
\caption{CT scan of an artery. The initial triangulation (left) that contains
  $63,468$ triangles has been split into $101$ surfaces that are
  parametrizable. Most of the cuts were done because of the large
  aspect ratio of the tubular domains. The uniform mesh on the right that contains $170,692$
  triangles has been generated by Gmsh in $22$ seconds, including
  IO's. \label{fig:artery}}
\end{figure}
\begin{figure}[!ht]
\begin{tabular}{cc}
\includegraphics[width=.45\linewidth]{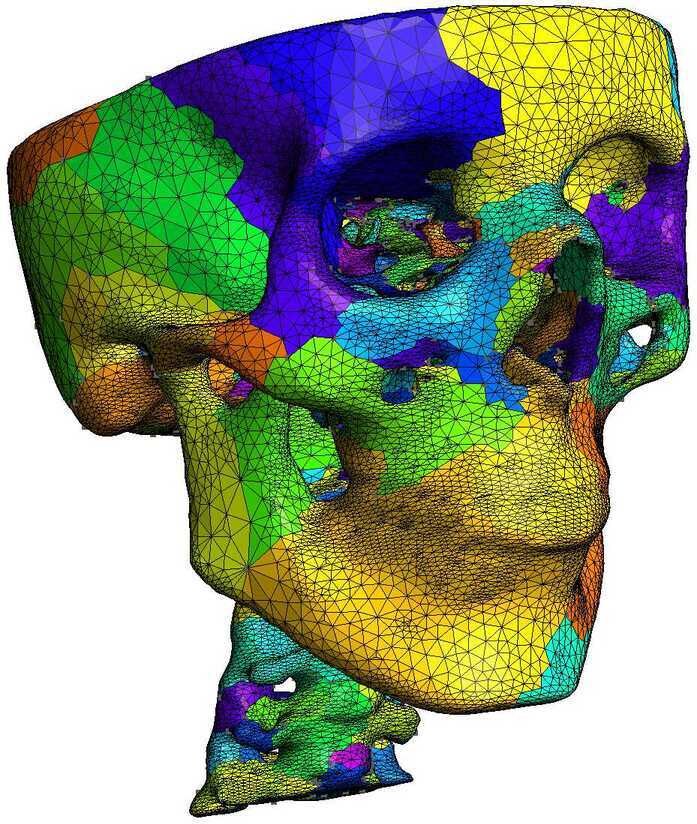} &  \includegraphics[width=.45\linewidth]{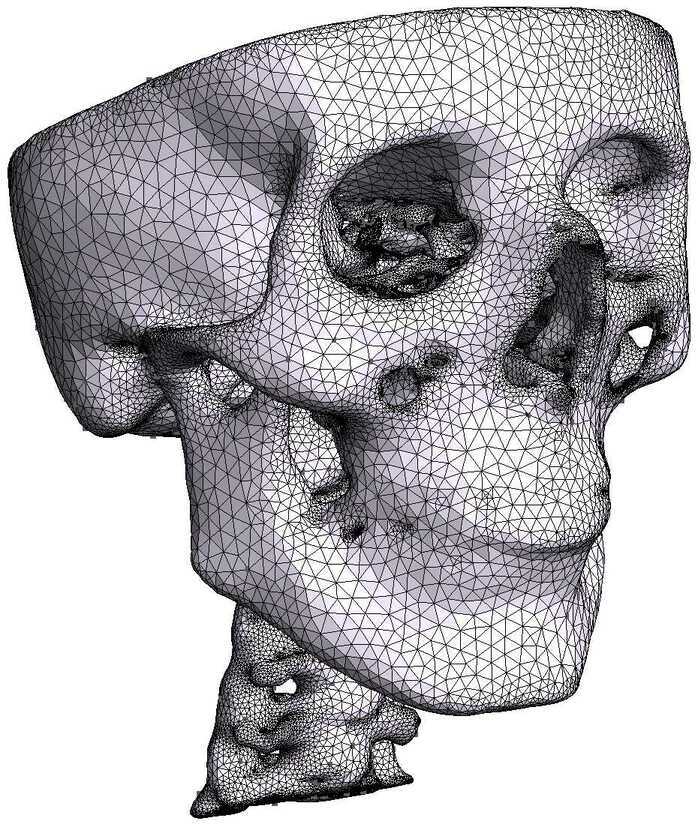}
\end{tabular}
\caption{Remeshing of a skull.  The initial triangulation (left) that contains
  $142,742$ triangles has been split into $715$ surfaces that are
  parametrizable. The mesh on the right is adapted to the surface
  curvature and  contains $323,988$
  triangles and has been generated by Gmsh in $58$ seconds, including
  IO's.\label{fig:skull}}
\end{figure}

\FloatBarrier
\section{Conclusion}
\label{sec:conclusion}

This paper has demonstrated the Gmsh's ability to remesh robustly poor quality
triangulations, for the purpose to run finite element analysis.  Gmsh's pipeline
essentially relies on the one-to-oneness of parametrization, where conformity is
not mandatory since a mesher has to deal with anisotropic meshes.  We have shown
that such a discrete parametrization is possible only if the corresponding
mapping orients all parametric triangles in the same way.

Based on the mean value theorem and assuming a linear approximation, we have
derived the well-known mean value coordinates.  We performed a convergence test
of the corresponding scheme: it does not discretize properly a Laplacian on a
structured mesh; otherwise, it has the expected convergence for a scheme that is
not symmetric.  We have proved that if homogeneous Neumann conditions are set along the
boundary of holes (within a triangulation), the mean value coordinates give
parametric holes whose boundary is convex.  Since it unnecessarily deforms the
parametrization, we gave an heuristic that fills the holes as they were circular
in order to produce better parametrizations.

With one simple but graphic example, we shown the effect of feature edge
detection on the atlas creation.  We have discussed how to improve the
parameterization of a coarse triangulation: the only way is to perform a longest
edge bisection before parametrization.  We have shown there is no Lagrange
$\mathcal{P}^2$ version of the mean value coordinates.  Finally, several
difficult examples were exhibited as a demonstration of the robustness of Gmsh's
pipeline.

\FloatBarrier
\section*{Acknowledgements}

The present study was carried out in the framework of the project ``Large Scale
Simulation of Waves in Complex Media'', which is funded by the Communauté
Française de Belgique under contract ARC WAVES 15/19-03.

\FloatBarrier
\bibliography{mybibfile}

\appendix
\FloatBarrier
\section{Derivation of the FEM scheme for harmonic mapping}
\label{appendixFEM}

\emph{Continuous} harmonic maps minimize the \emph{Dirichlet energy}
$$\int_{P_i} |\nabla \phi_i|^2 d\mathbf{x}$$
of the parametrization $\phi_i$ on the patch $P_i$.  In other words, it
minimizes the distortion between the patches and their planar representation.

It is possible to write a Laplacian as a finite difference scheme
$$
\nabla^2 f|_i \approx \sum_{j \in J(i)} \lambda_{ij}~(f_i-f_j)
$$
It is a linear approximation of a Laplace operator at a vertex $i$.  Indeed, the
Laplace operator corresponds to the Euler-Lagrange equations derived from the
Dirichlet energy
$$
\begin{array}{rcl}
\displaystyle{\int |\nabla f|^2 ~d\mathbf{x}} &\approx& \displaystyle{\int ||\sum_j f_j \nabla \phi_j||^2 ~d\mathbf{x}} \\
\displaystyle{\left.\dfrac{d}{df} \int |\nabla f|^2 ~d\mathbf{x}\right|_i} &\approx& 2 \displaystyle{\int \sum_j f_j \nabla \phi_j \cdot \nabla \phi_i~d\mathbf{x}}
\end{array}
$$
with $\phi_{\bullet}$ denoting the linear function shape associated to a node.
On a triangle $\mathcal{T}_{ijk}$, knowing that $\phi_i+\phi_j+\phi_k=1$ over
$\mathcal{T}_{ijk}$
$$
\int_{\mathcal{T}_{ijk}} f_i \nabla \phi_i  \cdot \nabla \underbrace{(1-\phi_j-\phi_k)}_{\phi_i} + f_j \nabla \phi_i \cdot \nabla \phi_j + f_k \nabla \phi_i \cdot \nabla \phi_k ~d\mathbf{x}
$$
Rewriting last relation with terms $(f_i-f_j)$ and $(f_i-f_k)$, we obtain
$$
\lambda_{ij} = - \int_{\mathcal{T}_{ijk}} \nabla \phi_i \cdot \nabla \phi_j ~ d\mathbf{x}
$$
which corresponds to the standard Galerkin finite element.

On triangle $\mathcal{T}_{ijk}$ (Fig. \ref{fig:thetafem})
$$
\nabla \phi_i \cdot \nabla \phi_j = |\nabla \phi_i| |\nabla \phi_j|~\cos(\pi-\theta_k)
$$
where
$$
|\nabla \phi_i| = \dfrac{1}{l_{ik}\sin(\theta_k)} \text{\hspace*{1cm}} |\nabla \phi_j| = \dfrac{1}{l_{jk}\sin(\theta_k)}
$$
with $l_{\bullet k}$ the length of edge $[\bullet k]$.
Knowing that $|\mathcal{T}_{ijk}|=\frac{1}{2} l_{ik} l_{jk} \sin(\theta_k)$
$$
- \int_{\mathcal{T}_{ijk}} \nabla \phi_i \cdot \nabla \phi_j ~ d\mathbf{x} = \dfrac{1}{2}\dfrac{\cos(\theta_k)}{\sin(\theta_k)}
$$
Adding the contribution of $\mathcal{T}_{ilj}$
$$
\lambda_{ij}^{\text{FEM}} := \dfrac{1}{2} \left( \dfrac{\cos(\theta_k)}{\sin(\theta_k)} + \dfrac{\cos(\theta_l)}{\sin(\theta_l)} \right).
$$

\FloatBarrier
\section{MVC difference scheme on a structured mesh is not a Laplacian}
\label{appendixMVCvsLaplace}

The MVC difference scheme relative to $f_i$ corresponds to
\begin{equation}\label{eqn:structuredMVC}
\dfrac{\sqrt{2}}{h} \left(4f_i - (f_{i1} + f_{i2} + f_{i4} + f_{i5})\right) + \dfrac{2-\sqrt{2}}{h} \left(2f_i - (f_{i3}+f_{i6})\right) = 0
\end{equation}

\begin{figure}[ht!]
\begin{center}
\includegraphics[width=.5\linewidth]{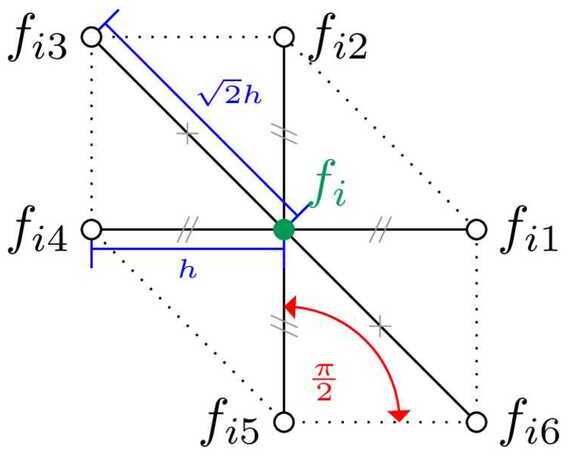}
\end{center}
\caption{Stencil within a structured mesh.}
\label{fig:stencilmvc}
\end{figure}

The first term of \eqref{eqn:structuredMVC} (without the coefficient)
corresponds to the well known linear combination of a centered finite difference
to approximate a Laplacian.  However, the second term does not approximate a
continous Laplacian.  Indeed, the Taylor expansion of $f_{i3}$ and $f_{f_i6}$ is
$$
\begin{array}{rcl}
  f_{i3} &=& f_i - \left.\dfrac{\partial f}{\partial x}\right|_i~h +
             \left.\dfrac{\partial f}{\partial y}\right|_i~h +
             \left.\dfrac{\partial^2 f}{\partial x^2}\right|_i~h^2 +
             \left.\dfrac{\partial^2 f}{\partial y^2}\right|_i~h^2 -
             \left.\dfrac{\partial^2 f}{\partial x \partial y}\right|_i~h^2 +
             \texttt{hot}\\
  f_{i6} &=& f_i + \left.\dfrac{\partial f}{\partial x}\right|_i~h -
             \left.\dfrac{\partial f}{\partial y}\right|_i~h +
             \left.\dfrac{\partial^2 f}{\partial x^2}\right|_i~h^2 +
             \left.\dfrac{\partial^2 f}{\partial y^2}\right|_i~h^2 -
             \left.\dfrac{\partial^2 f}{\partial x \partial y}\right|_i~h^2 +
             \texttt{hot}
\end{array}
$$
Hence,
$$
f_{i3} + f_{i6} - 2f_i = \overbrace{2\left.\dfrac{\partial^2 f}{\partial
      x^2}\right|_i~h^2 + 2\left.\dfrac{\partial^2 f}{\partial
      y^2}\right|_i~h^2}^{2h^2\left. \nabla^2 f\right|_i} -
2\left.\dfrac{\partial^2 f}{\partial x \partial y}\right|_i~h^2
$$
Because of that last term, the MVC scheme on a structured mesh such as
Fig. \ref{fig:stencilmvc} is not the approximation of a continuous Laplacian.

\begin{figure}[!ht]
\begin{center}
  \subfloat[FEM 
  computation.]{\includegraphics[width=.75\linewidth]{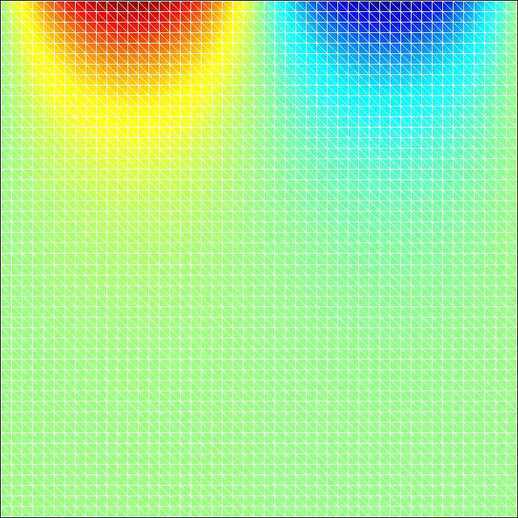}}\\
  \subfloat[MVC                                                       
  computation.]{\includegraphics[width=.75\linewidth]{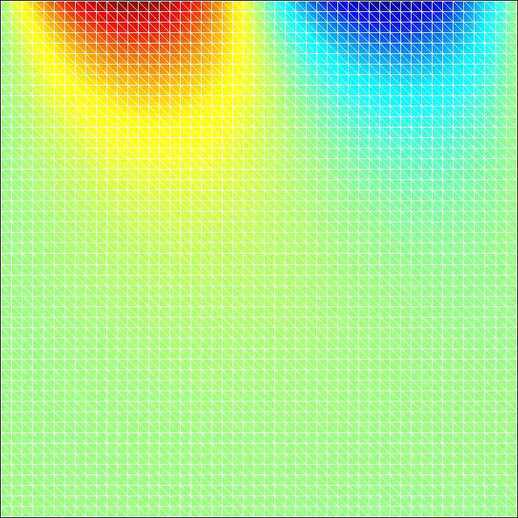}}
\caption{Approximations of $\nabla^2f=0$ on a structured mesh over a square $[0;1]\times[0;1]$.}
\label{fig:approxstructured}
\end{center}
\end{figure}
\end{document}